\documentclass[11pt]{article}
\usepackage{graphics,graphicx,color,amsmath,amssymb,threeparttable,wrapfig,mdframed}
\usepackage[margin=20truemm]{geometry}

\newcommand\arcsec{\mbox{$.\!\!^{\prime\prime}$}}

\title{\bf Support for fragile porous dust in a gravitationally self-regulated disk around IM Lup}

\author{
Takahiro Ueda$^{1,2*}$, 
Ryo Tazaki$^{3,4}$, 
Satoshi Okuzumi$^{5}$, 
Mario Flock$^{1}$ \& 
Prakruti Sudarshan$^{1}$
}

\date{\empty}

\begin{document}

\maketitle

\begin{itemize}
\item[] $^{1}$Planet and Star Formation Department, Max Planck Institute for Astronomy, K\"{o}nigstuhl 17, 69117 Heidelberg, Germany

\item[] $^{2}$Radio \& Geoastronomy Division, Harvard-Smithsonian Center for Astrophysics, 60 Garden Street, Cambridge, MA 02138, USA

\item[] $^{3}$Institute of Planetology and Astrophysics, Universit\'{e} Grenoble Alpes, CNRS, 38000 Grenoble, France

\item[] $^{4}$Astronomical Institute, Graduate School of Science Tohoku University, 6-3 Aramaki, Aoba-ku, Sendai, Miyagi, 980-8578, Japan

\item[] $^{5}$Department of Earth and Planetary Sciences, Tokyo Institute of Technology, 2–12–1 Ookayama, Meguro, Tokyo, 152-8551, Japan

\item[] $^{*}$Corresponding author, Email: takahiro.ueda@cfa.harvard.edu
\end{itemize}

\vspace{6pt}
\begin{abstract}
Protoplanetary disks, the birthplace of planets, are expected to be gravitationally unstable in their early phase of evolution. 
IM Lup, a well-known T-Tauri star, is surrounded by a protoplanetary disk with spiral arms likely caused by gravitational instability. 
The IM Lup disk has been observed using various methods, but developing a unified explanatory model is challenging.
Here we present a physical model of the IM Lup disk that offers a comprehensive explanation for diverse observations spanning from near-infrared to millimeter wavelengths.
Our findings underscore the importance of dust fragility in retaining the observed millimeter emission and reveal the preference for moderately porous dust to explain observed millimeter polarization.
We also find that the inner disk region is likely heated by gas accretion, providing a natural explanation for bright millimeter emission within 20 au. 
The actively heated inner region in the model casts a 100-au-scale shadow, aligning seamlessly with the near-infrared scattered light observation. 
The presence of accretion heating also supports the fragile dust scenario in which accretion efficiently heat the disk midplane. 
Due to the fragility of dust, it is unlikely that a potential embedded planet at 100 au formed via pebble accretion in a smooth disk, pointing to local dust enhancement boosting pebble accretion or alternative pathways such as outward migration or gravitational fragmentation.
\end{abstract}

\section*{Introduction}\label{sec1}

Planetary systems form within protoplanetary disks rotating around protostars.
The disks are expected to be unstable against their self-gravity during the early phase of their evolution (e.g.,  \cite{KL16}).
It is therefore of great importance to investigate the properties of the gravitationally unstable disk to understand the initiation of planet formation.

The unique feature of a gravitationally unstable disk is the presence of large-scale spiral arms, as observed in several objects \cite{Perez+16,Andrews+18}.
These spiral arms efficiently transport angular momentum and stabilize the disk against self-gravity, resulting in the gravitational self-regulation of the disk (e.g.,  \cite{Tsukamoto+15,KL16}). 
The spiral arms may fragment into gas blobs depending on their cooling efficiency, which lead to the direct formation of giant planets \cite{Helled+14}.

The protoplanetary disk around IM Lup is one of the most interesting and well-studied disks with spiral arms.
IM Lup is a T-Tauri star with a spectral type of K5 \cite{Alcala+17} and age of $\sim1$ Myr \cite{Avenhaus+18}.
The Disk Substructures at High Angular Resolution Project (DSHARP) with Atacama Large Millimeter/sub-millimeter Array (ALMA) revealed two-arm spiral structure extending out to $\sim100~{\rm au}$ in the IM Lup disk \cite{Andrews+18,Huang+18b}.
The DSHARP image also shows ring and gap-like structures just outside the outer edge of the spiral arms, which are potentially generated by an embedded giant planet.
The presence of a giant planet is also inferred from the kink structures in the CO line channel maps \cite{Pinte+20}.
These indicate that planet formation starts already at the gravitationally unstable disk phase.

The IM Lup disk has also been observed at near-infrared (NIR) wavelengths \cite{Avenhaus+18}.
The NIR polarized emission of the IM Lup disk exhibits extraordinary large disk with radius of $\sim400$ au with steeply flared disk surface \cite{Avenhaus+18}.
The NIR images also shows a shadow-like structure extending out to $\sim 100$ au, which is remarkable when the effect of stellar flux attenuation is corrected \cite{Avenhaus+18,Tazaki+23}.
The shadow potentially affects the physical/chemical condition of the disk by changing the radiation field within the disk (e.g.,  \cite{Ueda+19}).

In this article, we develop a theoretical model of dust evolution in the gravitationally self-regulated (GSR) disk around IM Lup. 
Our model is based on the assumption that the IM Lup disk is gravitationally self-regulated, i.e., marginally gravitationally unstable, and employs accretion heating and external radiation as heat sources. 
We demonstrate that our model comprehensively accounts for multiple observations of the IM Lup disk, spanning from near-infrared to millimeter wavelengths, and show that the dust in the IM Lup disk is likely fragile and moderately porous.

\section*{Overall picture of our model}
Here we describe overall picture of our model of the IM Lup disk.
A schematic overall picture of our model is shown in Fig. 1 and the detailed model description is provided in the Methods section.

\begin{figure*}[ht]
\centering
\includegraphics[width=1.0\textwidth]{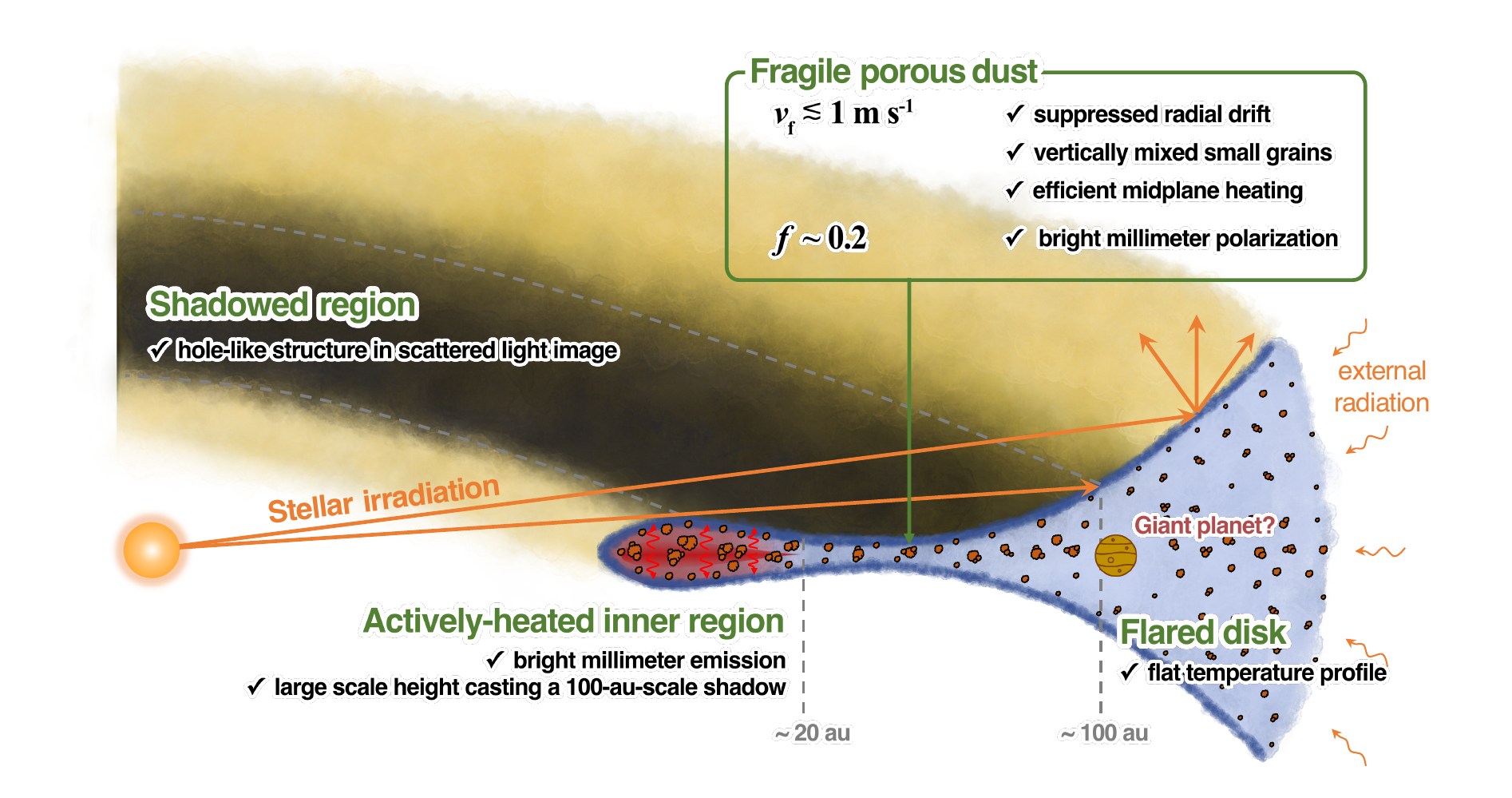}
\caption{
Schematic overall picture of our model of the IM Lup disk.
The midplane region inside $\sim20$ au is heated by gas accretion.
The inner region vertically puffs up due to its high temperature, casting a shadow extending out to $\sim100$ au.
The midplane temperature beyond 20 au is determined by external radiation because of the shadowing.
The flat temperature profile set by external radiation makes the disk surface flared strongly.
The fragile dust ($v_{\rm f}\lesssim1~{\rm m~s^{-1}}$) is preferred from both observations and temperature structure.
The moderate porosity is anticipated from the millimeter polarimetric observations.
}
\label{fig:schematic}
\end{figure*}

We construct a gas disk model (midplane temperature, gas surface density and turbulence strength) based on the assumption that the IM Lup disk is gravitationally self-regulated. 
This is motivated by the presence of spiral arms in the millimeter dust continuum emission  \cite{Andrews+18,Huang+18b}.
The midplane temperature is assumed to be determined by gas accretion at $\lesssim20$ au, while it is determined by external radiation outside 20 au.
Based on the gas disk model, we perform simulations of dust evolution with using DustPy  \cite{Stammler+22} with dust fragmentation velocity being a free parameter.
In addition. we consider two dust porosity models; fully compact dust (filling factor $f=1$) and moderately porous dust ($f=0.2$).
The obtained dust distribution is used to create synthetic images by using the so-called DSHARP opacity model  \cite{Birnstiel+18}. 
The synthetic images are compared with multiple observations to infer the dust properties.

The comparison of millimeter dust thermal emission (Stokes I) obtained from our models and ALMA observations suggests that the dust needs to be fragile; $v_{\rm f}\lesssim1~{\rm m~s^{-1}}$.
If $v_{\rm f}$ is as high as $10~{\rm m~s^{-1}}$, even a gravitationally self-regulated disk lacks sufficient dust to produce the observed millimeter emission.

In addition to the Stokes I emission, we also compare the polarized millimeter emission obtained from our models and ALMA observations.
The comparison suggests that the dust is moderately porous, with $f\sim0.2$.
If the dust is fully compact, reproducing the observed millimeter polarization is challenging, while moderately porous dust can produce self-scattering polarization with a broad dust-size range.
The presence of moderately porous dust has been also inferred for explaining the millimeter polarization of the HL Tau disk  \cite{Zhang+23}.
The moderate porosity is theoretically expected to be the consequence of hierarchical growth of dust aggregate  \cite{Dominik+16,Tanaka+23}.

The comparison of the NIR scattered-light image obtained from our models and the observation supports our hypothesis that the inner region of the IM Lup disk is likely heated by disk accretion.
The presence of accretion heating naturally explains the steep increase in the millimeter intensity at $r\lesssim20$ au as well as the 100-au-scale shadow seen in the NIR scattered light image.
The flat temperature structure assisted by the shadow and external heating is responsible for the steeply flared disk surface.
The presence of accretion heating also supports the fragile dust scenario in which accretion energy can efficiently heat the disk midplane.

In the following sections, we show our fragile porous dust model reasonably accounts for the diverse observations of the IM Lup disk.

\section*{Comparison with observations}\label{sec:result}
We compare our model with observations to deduce the disk properties. 
We utilize the observations of (1) the millimeter dust thermal emission taken by the ALMA DSHARP and MAPS programs, (2) polarized millimeter emission and (3) polarized near-infrared emission observed by VLT/SPHERE.
In addition, we also infer the dust properties by comparing our temperature model (Eq. \ref{eq:temp}) with those obtained from thermal radiative transfer simulations.

\subsection*{Millimeter dust thermal emission} \label{sec:cont}

\begin{figure*}[ht]
\centering
\includegraphics[width=1.0\textwidth]{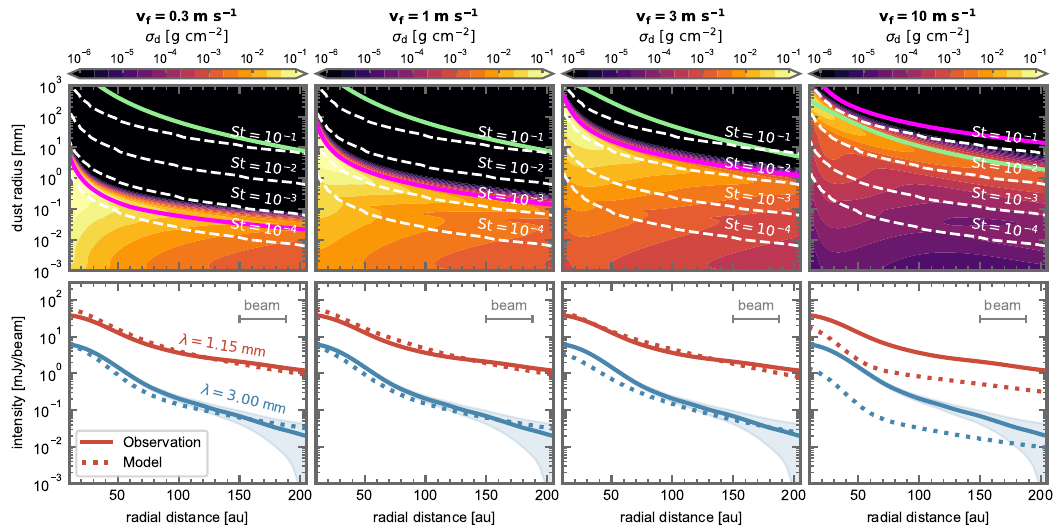}
\caption{
Dust distribution and radial intensity profile obtained from our simulations.
The critical fragmentation velocity is assume to be 0.3, 1, 3 and 10 ${\rm m~s^{-1}}$ from left to right.
The dust filling factor $f$ is set to 0.2 ($p=0.8$). 
Top: Dust-size distribution. The magenta and green lines in the top panels denote the fragmentation- and drift-regulated dust size, respectively \cite{Birnstiel+12}.
Bottom: Comparison between intensity profiles obtained from our model and observations. The solid lines represent the observed radial intensity profile with $1\sigma$ error denoted by transparent area obtained from ALMA MAPS program \cite{Sierra+21}.
}
\label{fig:fiducial}
\end{figure*}

We start the comparison with the millimeter dust thermal emission.
Fig. \ref{fig:fiducial} compares our models with porous ($f=0.2$) dust and the ALMA MAPS observations \cite{Sierra+21}.
The MAPS observations are at $\lambda=1.15$ (ALMA Band 6) and 3.00 mm (Band 3) and have angular resolution of 0$\arcsec$24, corresponding to the spatial resolution of $\sim38$ au.

Our porous dust model reasonably reproduces the observed intensity at both $\lambda=1.15$ and 3.00 mm when $v_{\rm f}\lesssim1~{\rm m~s^{-1}}$, while the model starts to deviate from the observations above $v_{\rm f}=3~{\rm m~s^{-1}}$. 
The model with $v_{\rm f}\lesssim1~{\rm m~s^{-1}}$ aligns with the observations within a relative difference of 70\% up to 200 au, while it is $\sim130\%$ for $v_{\rm f}=3~{\rm m~s^{-1}}$ and $\sim1000\%$ for $v_{\rm f}=10~{\rm m~s^{-1}}$.
This behavior can be understood as follows.
In our model, the maximum dust size is regulated by fragmentation when $v_{\rm f}\lesssim3~{\rm m~s^{-1}}$, while it is controlled by radial drift when $v_{\rm f}=10~{\rm m~s^{-1}}$.
The radial drift timescale of the mass-weighted average dust particle at 20 au is $\sim10^{7}$, $\sim10^{6}$, $\sim10^{5}$ and $\sim4\times10^{4}$ yr at $v_{\rm f}=0.3$, 1, 3 and $10~{\rm m~s^{-1}}$, respectively, and is increasing with radial distance.
Therefore, the dust radial drift has no significant impact on the dust surface density at $\gtrsim20$ au for $v_{\rm f}\lesssim1~{\rm m~s^{-1}}$.
For $v_{\rm f}=3~{\rm m~s^{-1}}$, the radial drift is still not significant at $\gtrsim50$ au, but becomes noticeable at $\lesssim50$ au.
The radial drift reduces the dust surface density at entire region of the disk when $v_{\rm f}=10~{\rm m~s^{-1}}$.
The dust grains grow beyond 1 cm at $\lesssim50$ au for $v_{\rm f}=3~{\rm m~s^{-1}}$ and at $\lesssim100$ au for $v_{\rm f}=100~{\rm m~s^{-1}}$.
At millimeter wavelengths, these large grains have smaller opacity compared to smaller grains, which also makes the disk fainter.

We found similar intensity profile for the compact dust model with $v_{\rm f}\lesssim 1~{\rm m~s^{-1}}$ as shown in Supplementary Fig. \ref{fig:comp_compact}. 
This is because the compact and porous dust have similar absorption opacity when dust size is sufficiently small ($\lesssim100~{\rm \mu m}$, see Supplementary Fig. \ref{fig:opac}).
However, the compact dust model with $v_{\rm f}=3~{\rm m~s^{-1}}$ exhibits a bump-like structure in the intensity profile due to the opacity enhancement at $\lambda \sim 2\pi a_{\rm max}$, which is not seen in the porous dust models and observations.
Over all, our models with $v_{\rm f}\lesssim1~{\rm m~s^{-1}}$ are consistent with the dust continuum observations taken by the ALMA MAPS program, regardless of the porosity.
The necessity of the dust fragility corresponding to $v_{\rm f}\lesssim1~{\rm m~s^{-1}}$ aligns with the prediction derived from the parametric fitting of the ALMA observations \cite{Jiang+24}.
We note that it is unlikely for the IM Lup disk to be significantly more massive than our model, given that our model is massive enough to trigger gravitational instability.

\begin{figure}[ht]
\centering
\includegraphics[width=1\textwidth]{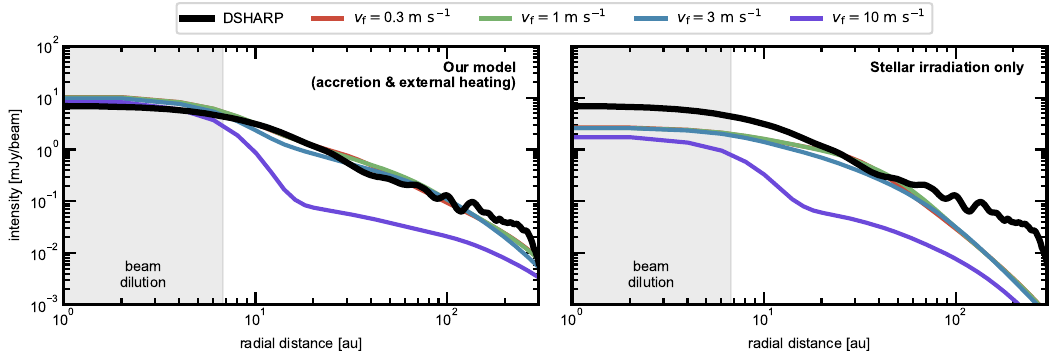}
\caption{
Comparison of the 1.25 mm intensity profile obtained from our models and the DSHARP observation.
The red, green, blue and purple solid lines denote the model with $v_{\rm f}=0.3$, 1, 3 and 10 ${\rm m~s^{-1}}$, respectively, while the black solid line denotes the observed value obtained from DSHARP observation \cite{Andrews+18}.
The left panel shows our standard model in which the disk temperature is determined by accretion heating and external irradiation, while the right panel shows the model based on the temperature model in which the heating is solely by stellar irradiation (Eq. \ref{eq:Sierra}).
}
\label{fig:dsharp}
\end{figure}

Fig. \ref{fig:dsharp} compares our porous dust models with the higher angular resolution observation ($\sim0\arcsec044$; $\sim 7$ au) taken by the ALMA DSHARP program ($\lambda=1.25$ mm) \cite{Andrews+18}.
For reference, we also show the model assuming passively irradiated disk in which the temperature is described by Eq. \eqref{eq:Sierra}.
Our model reasonably explains the overall trend of the observed intensity profile when $v_{\rm f}\lesssim3~{\rm m~s^{-1}}$.
In the outer region ($\gtrsim20$ au), the disk temperature is dominated by external heating ($T_{\rm ext}=15$ K) in our model, while it is down to a few K in the passive disk model.
The model incorporating external heating better aligns with the observation compared to the model considering only stellar radiation at $\gtrsim30$ au.
This is because higher temperature not only directly enhances thermal emission but also increases the surface density; a hotter disk can contain more mass while avoiding gravitational instability (Eq. \ref{eq:sigg}).

Another key point is that our model incorporating accretion heating well explains the steep increase in the intensity at $\lesssim20$ au, for which the model without accretion heating fails to explain.
At $\lesssim20$ au, the disk is fully optically thick at $\lambda=1.25$ mm when $v_{\rm f}\lesssim1~{\rm m~s^{-1}}$ (see Supplementary Fig. \ref{fig:tau}), and thus, the insufficient emission of the disk without accretion heating reflects the disk temperature that is not high enough.
Accretion heating contributes to the millimeter emission only at $\lesssim20$ au in our model, however, it has significant impact on the near-infrared emission in the outer region as shown later.

\subsection*{Millimeter polarized emission} \label{sec:pol}
Our models with $v_{\rm f}\lesssim 1~{\rm m~s^{-1}}$ reasonably explain the dust Stokes I emission observed at multiple ALMA wavelenths, but the porosity is not well constrained from the Stokes I emission alone.
Here, we compare our models with the ALMA polarimetric observations, which provide additional constraints on the the dust properties.

\begin{figure*}[ht]
\centering
\includegraphics[width=1.0\textwidth]{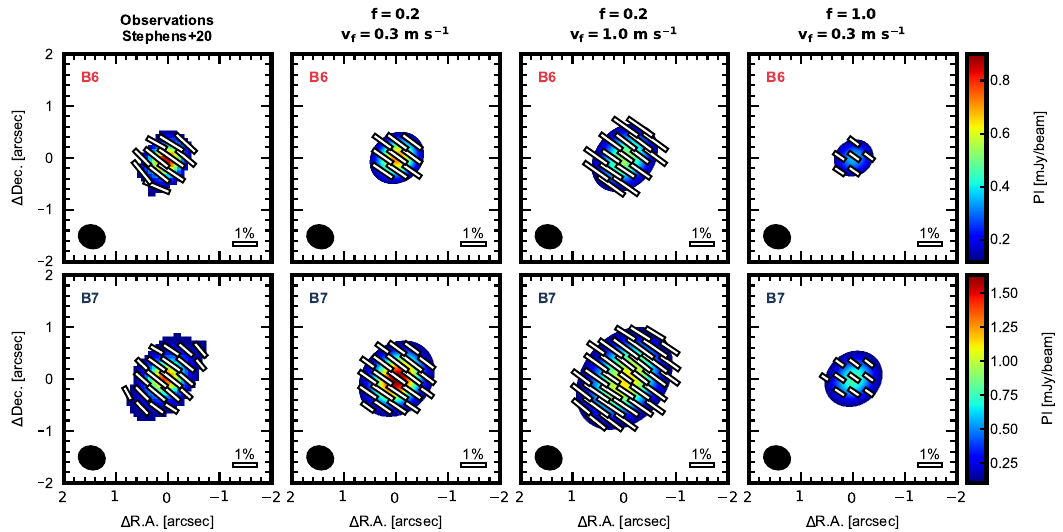}
\caption{
Polarized intensity along with the polarization vector obtained from the observations and our models.
The leftmost column shows the observed polarization \cite{Stephens+20} (see also  \cite{Hull+18}), whereas the other columns show the polarization images obtained from our porous dust model with $v_{\rm f}=0.3$ (second left) and $1~{\rm m~s^{-1}}$ (second right), and from compact dust model with $v_{\rm f}=0.3~{\rm m~s^{-1}}$ (rightmost).
The polarized intensity is shown only for the region where the polarized intensity is above the detection limit of the observations.
The length of polarization vector is scaled by the polarization fraction and 1\% polarization is denoted at bottom right in each panel.
The simulated intensity is smoothed with the observing beam size (0$\arcsec$51 $\times$ 0$\arcsec$44) denoted at bottom left in each panel.
}
\label{fig:pol}
\end{figure*}

The ALMA polarization observations toward IM Lup have been taken at Bands 6 \cite{Stephens+20} and 7 \cite{Hull+18}.
These observations clearly detected scattering-induced polarization, where the polarization vector is parallel to the disk minor-axis, with a polarization fraction of $\sim1$\% at central region ($\lesssim100$ au) of the disk.
Fig. \ref{fig:pol} compares the polarization morphology at ALMA Bands 6 and 7 obtained from our models with the observations.
The full suite of polarization images obtained from our models is shown in the Supplementary Fig. \ref{fig:polarization_all}).
Our porous dust models with $v_{\rm f}\lesssim 1~{\rm m~s^{-1}}$ show $\sim1$\% scattering-induced polarization at ALMA Bands 6 and 7.
More qualitatively, for $v_{\rm f}=0.3$, 1, 3 and $10~{\rm m~s^{-1}}$, our porous dust model predicts the polarization fraction of 0.85, 0.58, 0.66 and 0.40\% at Band 7 at the disk center, while the compact dust model shows only 0.42, 0.13, 0.10 and 0.17\% polarization, respectively.
As the measured central polarization fraction is 1\% at Band 7, the porous model with smaller $v_{\rm f}$ is more consistent with the observations.
In our porous dust model, detectable polarized emission comes from a wider region but the peak polarized intensity decreases as $v_{\rm f}$ increases from 0.3 to 3 ${\rm m~s^{-1}}$.
If $v_{\rm f}=10~{\rm m~s^{-1}}$, no significant polarization is seen in our model.
In our porous dust model, the polarization efficiency is maximized at $a_{\rm max}\sim 1~{\rm mm}$.
The maximum dust size exceeds $\sim1$ mm across most of the disk region when $v_{\rm f}=10~{\rm m~s^{-1}}$.

In contrast to the porous case, the compact dust models predicts significantly fainter polarized emission regardless of $v_{\rm f}$.
This is because compact dust produces scattering-induced polarization at a narrow dust-size range, whereas porous dust has broader dust-size range to produce polarization (see Supplementary Fig. \ref{fig:opac}) \cite{Tazaki+19,Zhang+23}.
The narrow dust-size range to produce polarization creates ring-like structure in the polarization image, which is noticeable in the compact dust model with $v_{\rm f}=1~{\rm m~s^{-1}}$.
This is because, in a standard disk model where the dust size increases toward the central star, the condition for maximizing polarization efficiency is achieved at a specific radial annulus where $a_{\rm max}$ is comparable to the observing wavelength \cite{OT19}.

Based on these results, we conclude that the porous dust model with $v_{\rm f}\lesssim 1~{\rm m~s^{-1}}$ is the most consistent with the millimeter observations.

\subsection*{Infrared scattered light} \label{sec:NIR}
We demonstrated that our porous dust model gives good agreement with the multi-wavelength ALMA observations which is sensitive to large dust around the midplane.
In this section, we compare our models with the observation at NIR wavelength, which traces the disk surface structure determined by the distribution of small grains.

\begin{figure*}[t]
\centering
\includegraphics[width=1.0\textwidth]{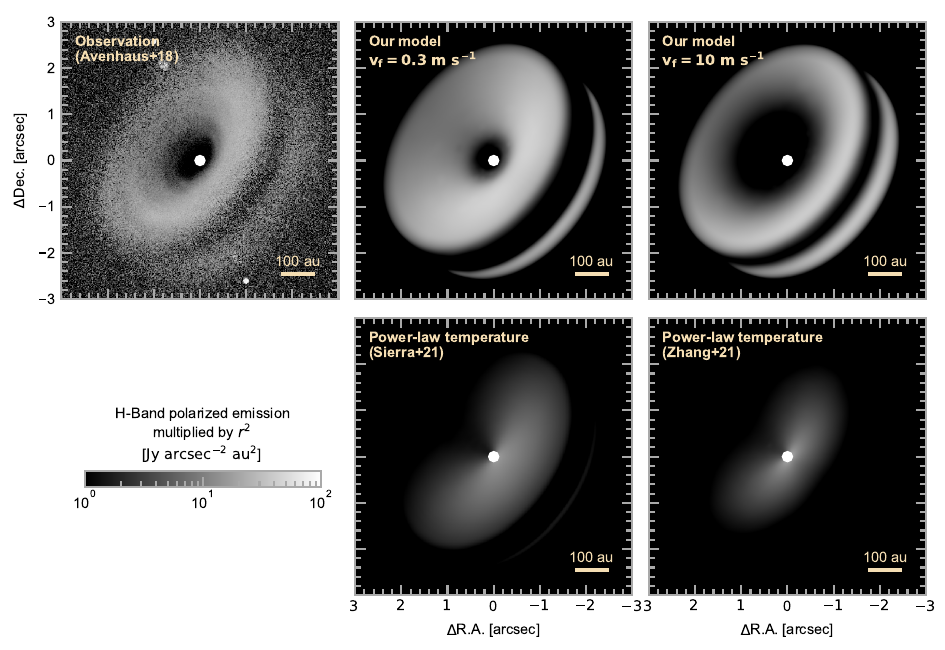}
\caption{
H-band ($1.6~{\rm \mu m}$) polarized scattered light multiplied by $r^{2}$. 
The $r^{2}$-scaling compensates for the stellar flux attenuation.
The VLT/SPHERE observation \cite{Avenhaus+18} is shown in the top left panel, while our porous dust models with $v_{\rm f}=0.3$ and 10 ${\rm m~s^{-1}}$ is shown in the top middle and top right panel, respectively.
For comparison, the bottom row shows the model assuming conventional power-law temperature profiles \cite{Sierra+21,Zhang+21} with $v_{\rm f}=0.3~{\rm m~s^{-1}}$.
}
\label{fig:NIR}
\end{figure*}

Fig. \ref{fig:NIR} compares the H-band ($\lambda=1.65~{\rm \mu m}$) polarized scattered light obtained from our models and the observation with VLT/SPHERE \cite{Avenhaus+18}.
The scattered light is multiplied by $r^{2}$ to compensate for the stellar flux attenuation.
The observed image clearly shows a hole-like structure extending out to $\sim100~{\rm au}$.
The hole-like structure is expected to be a shadow in which the disk surface receives no stellar irradiation.
Beyond the shadowed region, a flared disk surface is seen in the observed image.
Our model with fragile porous dust ($v_{\rm f}=0.3~{\rm m~s^{-1}}$)
reasonably reproduces the observed morphology of the polarized scattered light.
The shadowing is attributed to the inner region puffed up by accretion heating.
The flaring at $\gtrsim100$ au is due to the flat temperature profile induced by external irradiation.
In the case of $v_{\rm f}=10~{\rm m~s^{-1}}$, the amount of small grains is reduced compared to the case of $v_{\rm f}=0.3~{\rm m~s^{-1}}$. 
This depletion in small grains lowers the disk surface, resulting in much fainter scattered light for the model with $v_{\rm f}=10~{\rm m~s^{-1}}$ compared to both the model with $v_{\rm f}=0.3~{\rm m~s^{-1}}$ and the observed one.
If we assume conventional power-law temperature profiles \cite{Sierra+21,Zhang+21}, the shadow and steeply-flared surface structure are not seen in the synthetic images.
Furthermore, the power-law temperature model yields fainter polarized emission compared to the observation.

The ratio of the disk polarized flux to the total (star+disk) unpolarized flux is 0.65, 0.67, 0.47 and 0.26\% for $v_{\rm f}=0.3$, 1, 3 and $10~{\rm m~s^{-1}}$, respectively, while the observed value is 0.66$\pm$0.05\% \cite{Avenhaus+18}; fragile dust models are more consistent with the observation.
We note that the NIR polarization fraction depends on the detailed structure of dust aggregates and composition, which is not necessarily the same as those of the large dust observed by ALMA \cite{Tazaki+23}.

Based on this, we conclude that the our model incorporating accretion heating and fragile dust explains both the (sub-)millimeter observations and the NIR observation.

\subsection*{Validity of our temperature model} \label{sec:RADMC}
We attribute the steep increase in the millimeter intensity at $\lesssim20$ au and the NIR shadow to the high temperature induced by accretion heating.
As the temperature structure depends on the resulting dust distribution, we investigate the validity of our temperature model by performing thermal radiative transfer simulations based on the dust distribution obtained from our numerical simulations.

\begin{figure}[ht]
\centering
\includegraphics[width=1.0\textwidth]{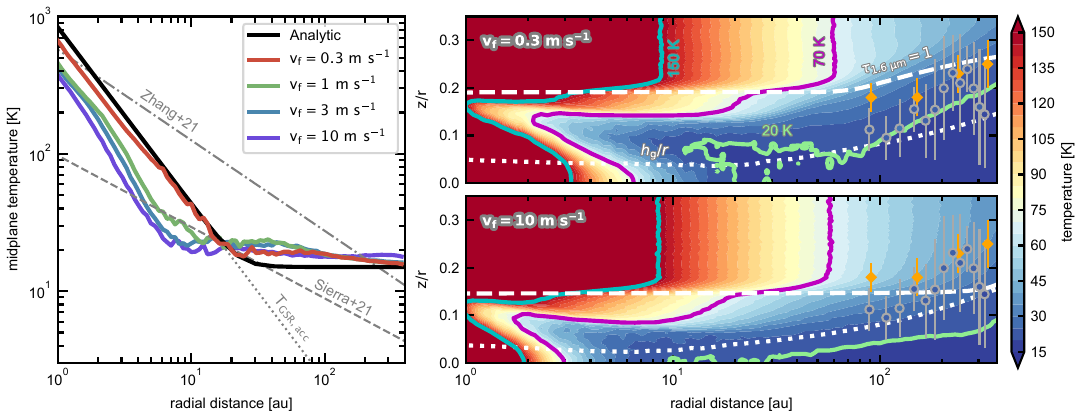}
\caption{
Temperature structure obtained from thermal radiative transfer simulations.
Left: midplane temperature obtained from the model with $v_{\rm f}=0.3$ (red), 1 (green), 3 (blue) and 10 ${\rm m~s^{-1}}$ (purple).
The black solid line denotes the analytical temperature used in our model.
The gray dotted, dashed and dash-dotted lines denote the temperature profile determined by accretion heating, temperature profile from  \cite{Sierra+21} and  \cite{Zhang+21}, respectively.
Right: two-dimensional temperature map of the model with $v_{\rm f}=0.3$ (top) and 10 ${\rm m~s^{-1}}$ (bottom).
The green, magenta and cyan curves denote the contour of 20, 70 and 150 K, respectively.
The white dotted and dashed lines denote the gas disk scale height and the NIR surface at which radial optical depth at $1.6~{\rm \mu m}$ reaches unity, respectively
The circles show the observed brightness temperature of the $^{13}{\rm CO}$ $J=$ 2--1 line emission \cite{Law+21}. The same color scale is applied for the points and the map. 
The brightness temperature is radially binned with uniform separation of 20 au and the errorbar represents $1\sigma$ uncertainty.
The orange diamonds show the measured height of ring-like structures seen in the NIR image \cite{Avenhaus+18} and the errorbar represents $1\sigma$ uncertainty.
}
\label{fig:temp}
\end{figure}

Fig. \ref{fig:temp} compares our assumed analytic temperature model (Eq. \ref{eq:temp}) with the temperature obtained from thermal radiative transfer simulations.
We see that the our analytic temperature model is in good agreement with that obtained from the radiative transfer simulation with $v_{\rm f}=0.3~{\rm m~s^{-1}}$.
The temperature at inner region ($\lesssim20$ au) decreases as $v_{\rm f}$ increases.
This is because larger $v_{\rm f}$ allow dust to grow larger which has smaller opacity.
The smaller opacity, with additional effect of surface density reduction by radial drift, makes the disk optically thinner in the vertical direction, leading to efficient radiative cooling.
The effect of accretion heating is also seen in the two-dimensional temperature map (Fig. \ref{fig:temp} right); the temperature increases toward the midplane at $\lesssim10$ au when $v_{\rm f}=0.3~{\rm m~s^{-1}}$, while the model with $v_{\rm f}=10~{\rm m~s^{-1}}$ shows the midplane heating at $\lesssim4$ au.
As shown in Supplementary Fig. \ref{fig:disk}, the bright emission at $\sim20$ au requires the dust temperature similar to our analytical model.

The two-dimensional temperature map is also compared with the brightness temperature of the $^{13}{\rm CO}$ $J=$ 2--1 line emission measured by  \cite{Law+21}. 
Interestingly, even though the $^{13}{\rm CO}$ brightness temperature traces the layer above one gas scale height, rather than the midplane, our model with $v_{\rm f}=0.3~{\rm m~s^{-1}}$, in which the stellar irradiation is absorbed well above the gas scale height, is more consistent with the measured brightness temperature.
Based on these, therefore, fragile dust is favored in terms of not only the disk observations but also the temperature structure.

\section*{Discussion}\label{sec:discussion}

\subsection*{Planet formation in the IM Lup disk}
We demonstrated that our model of the gravitationally self-regulated disk reasonably explains the multiple observations of the IM Lup disk.
The ALMA DSHARP program has shown that the IM Lup disk has gap and ring-like structures around $\sim100$ au (Fig. \ref{fig:dsharp}) \cite{Huang+18}.
The ALMA observations of molecular line emission also show kink structures in the CO line channel maps \cite{Pinte+20}.
These signatures are potentially attributed to an embedded planet with few Jupiter mass \cite{Verrios+22}.
In this section, we discuss the planet formation in the IM Lup disk based on our model.

The possibility of local dust accumulation if pressure trap is at play can be evaluated by comparing the Stokes number with the turbulence strength; the radial and vertical dust accumulation require ${\rm St_{\rm mass}} \gg \alpha_{\rm turb}$ \cite{Dullemond+18,Ueda+19}.
The mass-weighted averaged Stokes number obtained from our simulations is shown in Supplementary Fig. \ref{fig:pebble} (left).
For the fragile dust with $v_{\rm f}\lesssim1~{\rm m~s^{-1}}$, the mass-weighted Stokes number is lower than $\alpha_{\rm turb}$ beyond a few tens of au, which is unfavorable for planetesimal formation by local dust accumulation.
On the other hand, the inner region is more favorable for planetesimal formation; the mass-weighted Stokes number is larger than $\alpha_{\rm turb}$ at $\lesssim10$ au even for $v_{\rm f}=0.3~{\rm m~s^{-1}}$.

To investigate the feasibility of giant planet formation, we calculate the pebble accretion rate based on the dust size distribution obtained from our simulations.
We calculate the pebble accretion rate for each dust-size bin using the analytical formula \cite{LC18} and integrate it over the whole size distribution.
Supplementary Fig. \ref{fig:pebble} (right) shows the growth timescale of a planet accreting pebbles.
The growth timescale is defined as $t_{\rm grow}=M_{\rm p}/\dot{M}_{\rm p,acc}$, where $M_{\rm p}$ is the planetary mass and $\dot{M}_{\rm p,acc}$ is the pebble accretion rate.
We assume $M_{\rm p}=3M_{\oplus}$.

The pebble accretion rate is higher for larger $v_{\rm f}$ because of the combined effect of the more rapid radial drift and lower dust scale height.
We find that, at 100 au, the formation of a giant planet core may occur within 1 Myr only if $v_{\rm f}=10~{\rm m~s^{-1}}$. 
However, this high stickiness is incompatible with the observed millimeter and near-infrared emission of the IM Lup disk. 
If $v_{\rm f}\lesssim1~{\rm m~s^{-1}}$, the growth timescale is $\gtrsim10^{7}$ yr at 100 au, which is much longer than the age of IM Lup.
Therefore it is unlikely that a potential embedded planet at 100 au formed via pebble accretion in a smooth disk within the IM Lup's age.
This would underscore the importance of local dust enhancement \cite{Lau+22,JO+23} or alternative pathways such as outward migration \cite{Crida+09} or gravitational fragmentation \cite{Boss97}.
We note that in our model, at $\sim$100 au, the pebble accretion is in the so-called 3D regime as long as $M_{\rm p}\lesssim10M_{\oplus}$ in which the growth timescale is independent on $M_{\rm p}$ \cite{Ormel17}.

In contrast, the inner region is suitable not only for planetesimal formation but also for the planet formation by pebble accretion; the high ${\rm St}/\alpha$ enhances the pebble accretion efficiency and hence a core of giant planet can form within 1 Myr at $\lesssim4$ au even if dust is as fragile as $v_{\rm f}=0.3~{\rm m~s^{-1}}$.

\subsection*{Implications for accretion mechanisms}
Our model assumes that the disk accretion is driven solely by gravitational instability.
However, the other mechanisms such as the magnetorotational instability \cite{BH91} (MRI) or magnetohydrodynamical (MHD) disk winds \cite{Bai17} may be also responsible for the accretion.

If MRI or MHD wind generate accretion stress larger than that obtained from our model, the disk would be less massive than our model.
As demonstrated in this work, the disk is so bright at millimeter wavelengths that even a marginally gravitationally unstable disk can achieve the observed brightness only if $v_{\rm f}\lesssim1~{\rm m~s^{-1}}$.
This suggests that the accretion stress generated by the other mechanisms should not be much larger than gravitational-instability-induced accretion stress.

One possible scenario where MHD-driven accretion is suppressed is that the gas ionization is inhibited by the shadow.
We demonstrated that the actively-heated inner region casts a shadow extending out to $\sim100~{\rm au}$.
This shadowing effect may also prevent ultra-violet (UV) radiation from penetrating into the disk at $\lesssim100$ au.
The reduction of UV radiation would decrease the ionization degree of the disk gas, and consequently, weaken the activity of MRI or MHD disk wind.

The presence of accretion heating, if true, also suggests disk accretion induced by gravitational instability.
We attribute the steep increase in the millimeter intensity at the inner region and the shadowing effect to the internal heating induced by gas accretion.
If accretion is driven by the MHD wind, internal heating is suppressed because heat energy is released at the upper layer of the disk \cite{Mori+19}. 
We note, however, that MHD accretion can still cause enough internal heating  depending on the disk ionization state and opacity \cite{BL20,Kondo+23}.
Future studies incorporating disk self-gravity and magnetohydrodynamics will be crucial for deeper understanding of the evolution of the IM Lup disk.

\subsection*{Implications for disk chemistry}
We demonstrated that our shadowed disk model reasonably explains both the scattered light and millimeter dust continuum observations. 
We expect that the shadow also affects the gas disk's chemistry.
The shadowed inner region does not receive much stellar irradiation, while the flared outer region does.
The shadow reduces the disk temperature at the midplane, altering the radial location of snowlines of valatile species \cite{OU21,Notsu+22}.
Furthermore, the radial variation of the efficiency of stellar irradiation would affect the chemistry driven by UV radiation from the central star.
Indeed, ALMA observations of ${\rm HCO^{+}}$ in the IM Lup disk have shown that the UV-driven formation of ${\rm HCO^{+}}$ is likely enhanced in the outer region \cite{Seifert+21}.

The UV-driven chemistry also affects the composition of dust grains. 
In this work, we assume that the dust composition is uniform throughout the entire region of the disk. 
However, there might be a radial gradient of dust composition induced by UV-driven chemistry; the inner region is richer in organics, while the outer region, where UV-driven photodissociation of hydrocarbons may occur \cite{Mannella+96}, is richer in amorphous carbon .
Our model underestimates the millimeter intensity at $\gtrsim200$ au, even though the disk is massive enough for gravitational instability to be activated. 
If the dust in the outer region is rich in amorphous carbon, the absorption opacity can be higher than in our current model, potentially leading to better agreement with observations.
The presence of amorphous-carbon-rich dust is also inferred from the modeling of the near-infrared scattered light from outer region of the IM Lup disk \cite{Tazaki+23}.
The shadowing effect might affect UV-driven chemistry at the disk surface, not the midplane, as UV photons are rapidly attenuated at the surface \cite{Bergin+07}.

The ALMA MAPS program reported a depression in the ${\rm ^{13}CO}$ and ${\rm C^{18}O}$ line emission at $\lesssim20$ au in the IM Lup disk \cite{Bosman+21}, which may be attributed to the line absorption by dust with a total mass of $>500M_{\oplus}$ \cite{Bosman+23}.
In our fragile dust scenario, the total dust mass within 20 au is $\sim300M_{\oplus}$ and the disk is completely optically thick, even at $\lambda=3$ mm (Supplementary Fig. \ref{fig:tau}), which may affect the observed CO line flux.
We note that the dust pileup corresponding to the total mass of $>500M_{\oplus}$ requires efficient dust inflow from the outer region, which may be conflict with the bright millimeter emission at $>20$ au.
While a detailed analysis involving dust modeling combined with disk chemistry is beyond our scope, it would be crucial for comprehending molecular line observations and understanding the composition of both solid and gaseous materials in planet-forming disks.


\section*{Methods}

\subsection*{Disk model} \label{sec:diskmodel}

We construct a gas disk model assuming that the disk is gravitationally self-regulated (i.e., marginally unstable) \cite{Paczynski78,Gammie01,KL16,Tsukamoto+17,Yamamuro+23,XA23}. 
The stability against disk self-gravity is characterized by the Toomre's parameter \cite{Toomre1964}:
\begin{equation}
Q \equiv \frac{c_{\rm s}\kappa_{\rm epi}}{\pi G \Sigma_{\rm g}} \label{eq:Q}
\end{equation}
where $G$ is the gravitational constant, $\Sigma_{\rm g}$ is the gas surface density, $c_{\rm s}$ is the sound speed, $\kappa_{\rm epi}$ is the epicyclic freqeuncy.
For the Keplerian disk, $\kappa_{\rm epi}=\Omega_{\rm K}$ where $\Omega_{\rm K}=\sqrt{GM_{*}/r^{3}}$ is the Keplerian frequency with $M_{*}$ being the stellar mass.
The disk is gravitationally unstable if $Q\lesssim Q_{\rm crit}\approx1.4$ \cite{LD94}.

If $Q \lesssim Q_{\rm crit}$, accretion stress induced by gravitational instability decreases the gas surface density for the disk to be stable, and eventually $Q$ reaches $Q_{\rm crit}$.
Therefore, the gas surface density of the gravitationally self-regulated disk is given as
\begin{equation}
\Sigma_{\rm g, GSR} = \frac{c_{\rm s}\Omega_{\rm K}}{\pi G Q_{\rm crit}}. \label{eq:sigg}
\end{equation}
Assuming radially-constant mass accretion rate $\dot{M}$, the effective viscosity parameter for gas accretion $\alpha_{\rm acc,GSR}$ is obtained as 
\begin{equation}
\alpha_{\rm acc, GSR} = \frac{\dot{M}}{3\pi\Sigma_{\rm g, GSR}c_{\rm s}^{2}/\Omega_{\rm K}} = \frac{G\dot{M}Q_{\rm crit}}{3c_{\rm s}^{3}}. \label{eq:alp}
\end{equation}
Here we assume that the disk viscosity is described as $\nu=\alpha_{\rm acc,GSR}c_{\rm s}h_{\rm g}$ \cite{SS73} with $h_{\rm g}=c_{\rm s}/\Omega_{\rm K}$ being the gas scale height.
According to Equation \eqref{eq:sigg} and \eqref{eq:alp}, the radial profiles of gravitationally self-regulated disks are determined only by gas temperature once stellar properties are given. 
We assume $\dot{M}=10^{-7.9}M_{\odot}~{\rm yr^{-1}}$ and $M_{*}=1.1M_{\odot}$, which are typical estimates for IM Lup \cite{Alcala+17,Oberg+21}.

It is not trivial to estimate the disk temperature profile, but some models have been proposed for the IM Lup disk.
The ALMA large program Molecules with ALMA at Planet-forming Scales (MAPS) investigated the disk structure based on the parametric fitting of the spectral energy distribution \cite{Zhang+21}. 
They proposed the best fit power-law gas scale height: $h_{\rm g}=10(r/100~{\rm au})^{1.17}~{\rm au}$, corresponding to the midplane temperature of $T=42(r/50~{\rm au})^{-0.66}~{\rm K}$ assuming $h_{\rm g}=c_{\rm s}/\Omega_{\rm K}$.
However, a radiative transfer calculation based on this disk model with considering only stellar irradiation yields the midplane temperature of \cite{Sierra+21}
\begin{equation}
T_{\rm MAPS}=12.7(r/50~{\rm au})^{-0.52}~{\rm K},
\label{eq:Sierra}
\end{equation}
which is considerably lower than the midplane temperature inferred from the gas scale height.
One potential explanation for this inconsistency is that the disk temperature is influenced by heating process other than stellar irradiation, such as accretion heating.
The presence of accretion heating can be inferred from the brightness temperature of dust thermal emission; the brightness temperature obtained by the high angular resolution observation in the ALMA DSHARP Program \cite{Andrews+18} is higher than the midplane temperature obtained from the radiative transfer simulation considering only stellar irradiation (Eq. \ref{eq:Sierra}) at $<20$ au (see Supplementary Fig. \ref{fig:disk}).
This indicates that the actual dust temperature is higher than $T_{\rm MAPS}$ at least at $<20$ au.

In this work, therefore, we propose a novel temperature model for the IM Lup disk, which incorporates internal accretion heating and external radiation.
The disk temperature determined by accretion heating is estimated by assuming thermal equilibrium where the heating rate $\Lambda_{\rm heat}$ is balanced with the cooling $\Lambda_{\rm cool}$ \cite{XK21b}:
\begin{equation}
\Lambda_{\rm heat} = \frac{3\Omega_{\rm K}^{2}}{4\pi}\dot{M} \label{eq:heat}
\end{equation}
and
\begin{equation}
\Lambda_{\rm cool} = \frac{8\tau_{\rm z}}{1+0.875\tau_{\rm z}^{2}}\sigma T^{4}, \label{eq:cool}
\end{equation}
where $\tau_{\rm z}=\Sigma_{\rm g}\kappa_{\rm R}/2$ is the vertical optical depth for the disk thermal emission with $\kappa_{\rm R}$ being the Rosseland-mean opacity and $\sigma$ is the Stefan-Boltzmann constant.
Equating \eqref{eq:heat} and \eqref{eq:cool} with using \eqref{eq:sigg}, the disk temperature determined by accretion heating is obtained as
\begin{equation}
T_{\rm GSR,acc} = \left( \frac{k_{\rm B}}{m_{\rm g}} \right)^{1/7} \left( \frac{21\dot{M}\Omega_{\rm K}^{3}\kappa_{\rm R}}{512\pi^{2}\sigma G Q_{\rm crit}} \right)^{2/7},
\label{eq:T_acc}
\end{equation}
where $m_{\rm g}$ is the mean mass of gas molecules and $k_{\rm B}$ is the Boltzmann constant.
Equation \eqref{eq:T_acc} gives the radial dependence of $T_{\rm GSR,acc} \propto \Omega_{\rm K}^{6/7}\propto r^{-9/7}$ with assuming $\kappa_{\rm R}$ being constant.
Interestingly, as shown in Fig. \ref{fig:disk}, the power-law index of $-9/7$ is in good agreement with the observed brightness temperature at ALMA Band 6 at $\lesssim30$ au. 
We assume that the overall temperature profile is determined as 
\begin{equation}
T^{4}=T_{\rm GSR,acc}^4+T_{\rm ext}^{4}, \label{eq:temp}
\end{equation}
where $T_{\rm ext}=15~{\rm K}$ is the temperature determined by external radiation.
The choice of $T_{\rm ext}=15~{\rm K}$ is motivated by the bright millimeter emission at $\gtrsim 100$ au and by the flared disk surface observed with VLT (see Fig. \ref{fig:dsharp} and \ref{fig:NIR}, and corresponding sections). 
Equation \eqref{eq:temp} ignores the stellar irradiation because steep increase in the temperature, and hence the gas scale height, toward the central region shields the stellar irradiation onto the outer region.
We set the Rosseland-mean opacity $\kappa_{\rm R}$ (per gas mass) to $0.5~{\rm cm^{2}~g^{-1}}$ such that the dust temperature aligns with the observed brightness temperature at $\sim10$--20 au.
The Rosseland-mean opacity depends on the dust properties such as size distribution, mass density and composition.

Our model of the gravitationally self-regulated disk around IM Lup is shown in Supplementary Fig. \ref{fig:disk}.
The gas surface density steeply increases toward the central region due to the steeply-increasing temperature ($\Sigma_{\rm g}\propto c_{\rm s}\Omega_{\rm K} \propto r^{-15/7}$).
We set the exponential cutoff radius of 385 au for the gas surface density \cite{Tazaki+23}.
The total gas mass within 385 au is $\sim0.44 M_{\odot}$ ($\sim0.4 M_{*}$), with roughly half of the mass, $\sim0.2 M_{\odot}$, contained within 20 au.
Although this total mass is $\sim2$-4 times higher than those commonly adopted, $\sim$0.1-0.2$M_{\odot}$ \cite{Cleeves+16,Zhang+21,Powell+22,Lodato+23,Martire+24}, our gas surface density model aligns with that of  \cite{Zhang+21} between 20 and 200 au.

Our theoretical model predicts $\alpha_{\rm acc}\sim2\times10^{-3}$ at $\gtrsim30$ au, which is in good agreement with the estimate by  \cite{Franceschi+23} in which the turbulence strength is evaluated from the vertical dust distribution inferred from VLTI/SPHERE and ALMA observations.
The recent observations of the molecular line broadening indicates strong turbulence corresponding to $\alpha\sim0.1$ at the upper layer of the disk \cite{Rosotti23,Paneque+24}. 
Our $\alpha_{\rm acc}$ represents the turbulence strength around the midplane, where most of the disk mass resides, suggesting a vertical gradient in turbulence strength \cite{Jiang+24}.
We note that the turbulence strength steeply decreases at $\lesssim20$ au and reaches $\sim10^{-5}$ at $\sim1$ au.
It is uncertain if gravito-turbulence can achieve the low $\alpha_{\rm acc}$ as nurimerical simulations have suggested $\alpha_{\rm acc}\sim0.01$-0.01 \cite{Gammie01}.
One possible explanation is that the inefficient disk cooling due to high optical depth (cooling timescale $\sim3\times10^{4}/\Omega_{\rm K}$ at 3 au) stabilizes the disk against the instability \cite{Rice+11,Paardekooper12,Bethune+21}.

\subsection*{Simulations of dust evolution} \label{sec:dustpy}
We perform simulations of dust evolution based on the analytical disk model using Dustpy \cite{Stammler+22}. 
The dust size distribution is crucial for comparing the model to multiple observations spanning from near-infrared to millimeter wavelengths. 
One key parameter in the dust evolution is the critical fragmentation velocity $v_{\rm f}$, which represents the velocity threshold above which colliding dust particles fragment into smaller particles \cite{Birnstiel+11}.
If the collisional velocity $v_{\rm coll}$ is higher than $v_{\rm f}$, the maximum dust size is regulated so that $v_{\rm f}\sim v_{\rm coll}$.
On the other hand, if $v_{\rm f} \gg v_{\rm coll}$, the colliding particles merge into larger and the maximum dust size is regulated by the dust radial drift \cite{Birnstiel+11}.
The collisional velocity is mostly regulated by the turbulent motion of dust induced by gas turbulence; $v_{\rm coll}\sim\sqrt{3\alpha_{\rm turb}{\rm St}}c_{\rm s}$ \cite{OC07} where we assume $\alpha_{\rm turb}=\alpha_{\rm acc}$, even though we also consider other contributions (e.g., brownian motion and differentical drift velocities).

The actual value of $v_{\rm f}$ depends on the detailed dust properties such as composition and monomer size and is highly uncertain \cite{Beitz+11,Wada+13,Arakawa+16,Okuzumi+16,MW19}.
For instance, at the outer region of the disk where temperature is below $\sim70$ K, the ${\rm CO_{2}}$ gas is expected to be frozen onto dust, which makes dust less sticky compared to pure ${\rm H_{2}O}$ ice \cite{Musiolik+16a,Musiolik+16b,OT19}.
We vary $v_{\rm f}$ as a free parameter to see how observables depend on it.

We consider two models for the dust porosity $p$; 0 and 0.8.
The former corresponds to the completely compact dust, while the latter corresponds to moderately porous dust.
The dust material density is set as $\rho_{\rm m}=1.675~{\rm g~cm^{-3}}$ and the dust internal density $\rho_{\rm s}$ is set as $\rho_{\rm s}=f\rho_{\rm m}$ where $f=1-p$ is the dust filling factor.
We do not consider extremely porous dust ($f\ll0.1$) because extremely porous dust is unlikely to account for the presence of scattering-induced polarization at millimeter wavelengths \cite{Tazaki+19} (Supplementary Fig. \ref{fig:opac}).
On the other hand, moderate porosity is favorable for explaining the ubiquity of the scattering-induced polarization \cite{Zhang+23}.
Furthermore, moderate porosity is motivated by theoretical works considering hierarchical dust growth in which small grains fill voids in large dust up \cite{Dominik+16,Tanaka+23}.

The simulated domain ranges from 1 to 1000 au in the radial direction, with logarithmically separated into 100 bins.
The dust-mass domain is from $10^{-12}$ to $10^{8}~{\rm g}$, with logarithmically divided into 10 bins per decade.
The simulations are performed up to $1~{\rm Myr}$, corresponding to the age of IM Lup.

\subsection*{Opacity model} \label{sec:opac}
We conduct radiative transfer simulations using RADMC-3D \cite{RADMC} to generate synthetic images of our model for comparing with observations. 
In this section, we describe our opacity model. 

The dust composition is based on the DSHARP model \cite{Birnstiel+18} which is a mixture of water ice \cite{WB08} (20\% in mass fraction), astronomical silicates \cite{Draine03} (33\%), troilite \cite{HS96} (7\%) and refractory organics as a carbonaceous material (40\%).
Based on the DSHARP composition, dust opacities are calculated with Optool \cite{Optool}.
We first generate the optical constant of the mixture using the Bruggeman rule. 
The mixture is then combined with vacuum, again utilizing the Bruggeman rule to account for porosity. 
It is worth noting that the vacuum component can be blended with other materials in a single step, but this approach results in a smaller absorption opacity compared to what is obtained through our method.
The smaller absorption opacity makes it challenging for our model to reproduce the bright emission observed by ALMA, despite the model being sufficiently massive to induce gravitational instability.
Therefore, with the opacity computed in the single step, one would need to assume that the dust contains more absorbing material such as amorphous carbon \cite{Zubko+96}.
In the context of dust evolution, however, porosity is likely due to the aggregate structure created by monomers that are the mixture of the dust materials.
Therefore, for the opacity calculation, it is likely more reasonable to treat the vacuum component as a component separated from the other dust materials.

We mimic the irregularity of dust shape (i.e., deviation from the perfect sphere) by using the distribution of hollow spheres method (DHS) with $f_{\rm max}=0.8$, where $f_{\rm max}$ describes the maximum fraction of the hollow \cite{Min+05}.
The inclusion of irregularity increases the absorption opacity compared to the perfect sphere model.
Furthermore, the irregularity acts as a small-scale substructure, causing scattering polarization even when the dust size exceeds the observing wavelength.

Opacities of our dust model is shown in Supplementary Fig. \ref{fig:opac}. 
The inclusion of porosity decreases the absorption opacity because small-scale substructures in porous dust suppresses the opacity enhancement at $\lambda\sim2\pi a_{\rm max}$ which is unique for the compact dust \cite{Kataoka+14}.
Supplementary Fig. \ref{fig:opac} (right) shows the product of the $90^{\circ}$-scattering probability $P$ and effective scattering albedo $\omega_{\rm eff}$, which describes the efficiency of self-scattering polarization \cite{Kataoka+15}.
The compact dust has a strong peak in the polarization efficiency at $a_{\rm max}\sim\lambda/2\pi$.
In contrast, porous dust has a broader peak in the polarization efficiency because small-scale substructures in large dust can produce scattering polarization.
We note that in our radiative transfer simulations, we do not assume the dust-size distribution but directly use the dust-size distribution obtained from our numerical simulations.

\subsection*{Imaging simulations} \label{sec:imaging}
Based on the dust distribution obtained from the DustPy simulations and opacity model described above, we perform Monte Carlo radiative transfer simulations to generate synthetic images of the model disk.
For the computational purpose, we regrid the dust size distribution obtained from the dust evolution simulations.
For the millimeter imaging, the dust size bins range from $10~{\rm \mu m}$ to $100$ cm, logarithmically divided into 125 bins. 
For the near-infrared imaging, the size bins range from $0.1~{\rm \mu m}$ to $1$ cm, logarithmically separated into 50 bins.

We assume that the vertical dust distribution is determined by the balance between turbulent mixing and settling and compute the dust scale height of each dust bin as \cite{Woitke+23}:
\begin{equation}
h_{\rm d}(a,z)=h_{\rm g} \left\{ 1 + \frac{\rm St}{\alpha_{\rm z}} \left( \frac{2h_{\rm g}^{2}}{z^{2}} \right)\left( \exp{\left( \frac{z^{2}}{2h_{\rm g}^{2}} \right)}-1 \right) \right\}^{-1/2},
\label{eq:settling}
\end{equation}
where ${\rm St}$ is the Stokes number of each dust bin at the midplane and $\alpha_{\rm z}=\alpha_{\rm acc}/3$ is the vertical mixing efficiency.
The prefactor 1/3 comes from the assumption of isotropic turbulence \cite{FP06}.
Equation \eqref{eq:settling} incorporates the dependence of the Stokes number on the vertical gas density profile. 
We conducted simulations using the settling prescription without considering the z-dependent St \cite{YL07} and confirmed that the effect of the z-dependent St does not impact our conclusion.

We use $10^{7}$ photon packages for the synthetic images of dust continuum emission at ALMA wavelengths.
For the ALMA polarization images and NIR images, we use $10^{9}$ and $10^{11}$ photon packages, respectively.
The full scattering is solved using scattering matrix.
The computational domain in the polar direction ranges from $-\pi/12$ to $\pi/12$ and is linearly separated into 64 bins for millimeter imaging, while it ranges from $-\pi/6$ to $\pi/6$ and is linearly separated into 64 bins for near-infrared imaging. 
The radial gridding is the same as that of the dust evolution simulation.
The disk inclination and position angle are set to $47.5^{\circ}$ and $144.5^{\circ}$, respectively \cite{Huang+18}. 
The obtained synthetic images are smoothed by a Gaussian function corresponding to the observing beam of the actual observation.

\subsection*{Thermal radiative transfer simulations} \label{sec:mctherm}
We also perform thermal radiative transfer simulations with RADMC-3D \cite{RADMC} to verify the validity of our temperature model.
We consider three components of the heating source: stellar irradiation, internal accretion heating, and external irradiation.
For accretion heating, we employ a conventional constant-$\alpha$ model where the specific heating rate is described by local viscosity; $q_{\rm acc}=(9/4)\alpha_{\rm acc}\rho_{\rm g}c_{\rm s}^{2}\Omega_{\rm K}$ with $\alpha_{\rm acc}$ being constant along the vertical direction.
The modified random walk method is applied for calculating the very optically thick regions.
The external radiation is assumed to be the black-body radiation with temperature of $15~{\rm K}$.
As the resultant temperature structure depends on the input density structure, which is in turn influenced by the temperature ($h_{\rm d}\propto h_{\rm g}\propto c_{\rm s}$), the thermal radiative transfer simulations are iteratively performed using the temperature structure obtain from the previous iteration step, as done in previous studies \cite{Ueda+19}. 

To reduce computational costs, the dust size distribution obtained from the dust evolution models is resampled into 11 bins: a dust size bin of  $a<10^{-3}$ mm and ten bins logarithmically separating the size ranging from $10^{-3}$ to $10^{2}$ mm.
We ignore the dust population larger than 10 cm as large dust does not contribute to the opacity.
We adopt $10^{7}$ photon packages for the thermal radiative transfer simulations. 
The scattering of photons is included with the Henyey-Greenstein approximation \cite{HG41}.

\vspace{12pt}
\noindent\textbf{Data availability}
The observational data used in this work are published in refs  \cite{Andrews+18,Avenhaus+18,Oberg+21,Hull+18,Stephens+20}. 
Due to the large size of the data files, the full physical data generated by the simulations presented in this work are available from the corresponding author upon request.

\vspace{12pt}
\noindent\textbf{Code availability} The numerical codes are  available from the corresponding author upon reasonable request.

\vspace{12pt}
\noindent\textbf{Acknowledgements} We acknowledge Ian Stephens, Christian Ginski, Anibal Sierra, Charles Law and Teresa Paneque-Carre\~{n}o and ALMA DSHARP and MAPS programs for providing us observational data used in this work. 
T.U. also thank Mizuki Ueda for creating schematic illustration shown in Fig. 1. Numerical computations were in part carried out on Small Parallel Computers at Center for Computational Astrophysics, National Astronomical Observatory of Japan, and on the Smithsonian High Performance Cluster (SI/HPC), Smithsonian Institution (https://doi.org/10.25572/SIHPC).

\vspace{12pt}
\noindent\textbf{Author contributions} The project was initiated through an informal conversation between T.U., R.T., and S.O. T.U. constructed the disk model with advice from F.M. and P.S. T.U. performed all the numerical simulations, with technical advice on the radiative transfer simulations provided by F.M. and P.S. R.T. and S.O. contributed to the opacity modeling. All authors provided comments used in editing the manuscript.

\vspace{12pt}
\noindent\textbf{Funding} T.U. acknowledges the support of the Deutsche Forschungsgemeinschaft (DFG, German Research Foundation) through the grant number 465962023 and the JSPS overseas research fellowship. R.T. acknowledges funding from the European Research Council (ERC) under the European Union's Horizon Europe research and innovation program (grant agreement No. 101053020, project Dust2Planets. M.F. acknowledges funding from the European Research Council (ERC) under the European Union’s Horizon 2020 research and innovation program (grant agreement No. 757957). P.S. acknowledges the support by the DFG through the grant number 495235860. This work is also supported by JSPS KAKENHI Grant Numbers JP23H01227, JP23K25923 and JP20H00182. 

\vspace{12pt}
\noindent\textbf{Correspondence} Correspondence and requests for materials
should be addressed to Takahiro Ueda.

\vspace{12pt}
\noindent\textbf{Competing Interests} The authors declare no competing interests.


\renewcommand{\refname}{References in the main text}

\renewcommand{\refname}{References in Methods}

\newpage
\section*{Supplementary information}
\setcounter{figure}{0}
\renewcommand{\figurename}{Supplementary Figure}
\begin{itemize}
\item{Supplementary Fig. \ref{fig:comp_compact}:} we show the dust size distribution and intensity profiles obtained from our compact dust models.
\item{Supplementary Fig. \ref{fig:opac}:} we show the absorption opacity and polarization efficiency at $\lambda=1.3$ mm of our dust model.
\item{Supplementary Fig. \ref{fig:tau}:} we show the dust surface density and vertical optical depths of our porous dust models.
\item{Supplementary Fig. \ref{fig:polarization_all}:} we present the complete set of polarization images obtained from our model.
\item{Supplementary Fig. \ref{fig:disk}:} we show our model of the IM Lup disk; midplane temperature, gas surface density and accretion stress parameter.
\item{Supplementary Fig. \ref{fig:pebble}:} we show the mass-weighted averaged Stokes number and core growth timescale via pebble accretion obtained from our numerical simulations.
\item{Parameter uncertainties:}
we briefly summarize the potential uncertainties in the parameters used in our model.
\end{itemize}

\begin{figure*}[ht]
\centering
\includegraphics[width=1.0\textwidth]{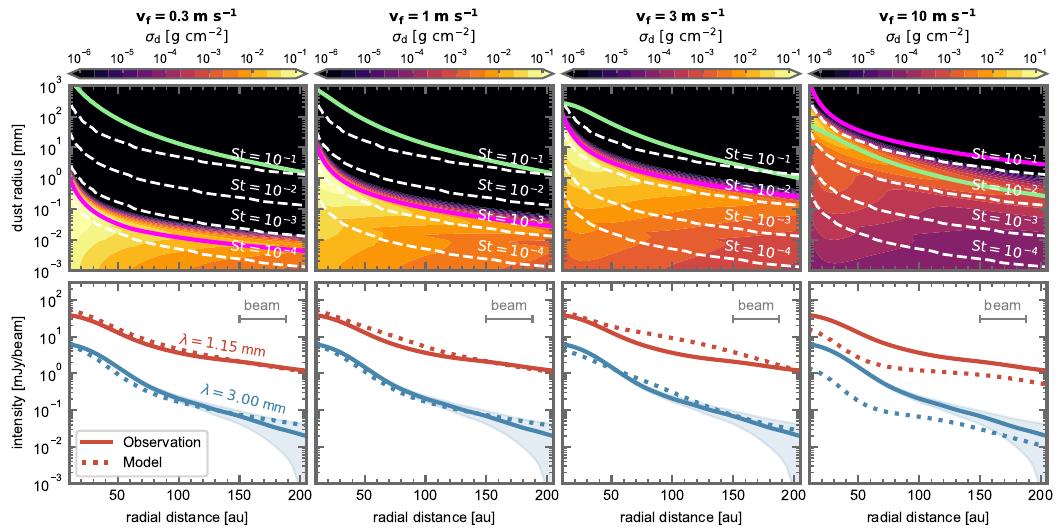}
\caption{
Dust distribution and radial intensity profile obtained from our simulations.
The critical fragmentation velocity is assume to be 0.3, 1, 3 and 10 ${\rm m~s^{-1}}$ from left to right.
The dust filling factor $f$ is set to 1 ($p=0$, i.e., fully compact dust). 
Top: Dust-size distribution. The magenta and green lines in the top panels denote the fragmentation- and drift-regulated dust size, respectively \cite{Birnstiel+12}.
Bottom: Comparison between intensity profiles obtained from our model and observations. The solid lines represent the observed radial intensity profile with $1\sigma$ error denoted by transparent area obtained from ALMA MAPS \cite{Sierra+21}.
}
\label{fig:comp_compact}
\end{figure*}

\begin{figure*}[ht]
\centering
\includegraphics[width=1.0\textwidth]{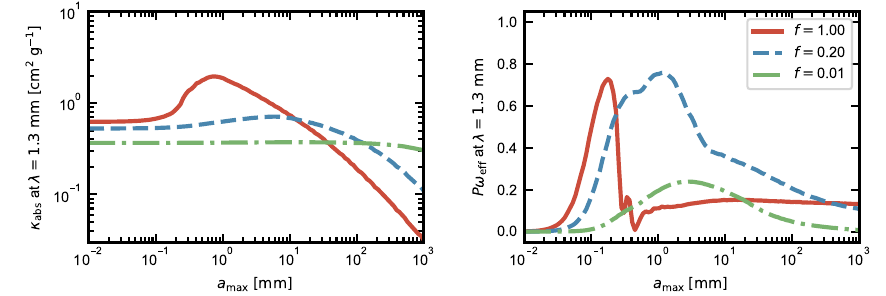}
\caption{
Opacities at $\lambda=1.3$ mm obtained from our dust model.
Left: absorption opacity.
Right: polarization efficiency which is described by the product of the degree of linear polarization at scattering angle of $90^{\circ}$ and effective albedo.
The red solid line and blue dashed line denote compact ($f=0$) and porous ($f=0.2$) dust model, respectively.
The green dash-dotted line denotes the dust opacity with $f=0.01$ for reference.
The dust size distribution is assumed to have a power-law index of $-3$.
}
\label{fig:opac}
\end{figure*}

\begin{figure*}[ht]
\centering
\includegraphics[width=1.0\textwidth]{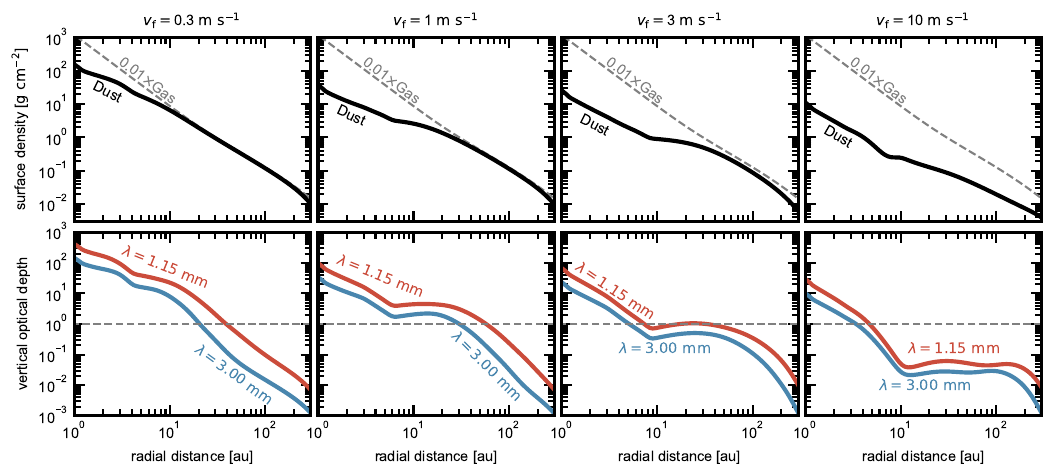}
\caption{
Surface density and vertical optical depths of our porous dust models.
Top: dust (black solid line) and gas (gray dashed line; multiplied by 0.01) surface densities.
Bottom: vertical optical depth at $\lambda=1.15$ (red) and 3.00 mm (blue).
The gray horizontal line denotes the optical depth of unity.
}
\label{fig:tau}
\end{figure*}

\begin{figure*}[ht]
\centering
\includegraphics[width=1.0\textwidth]{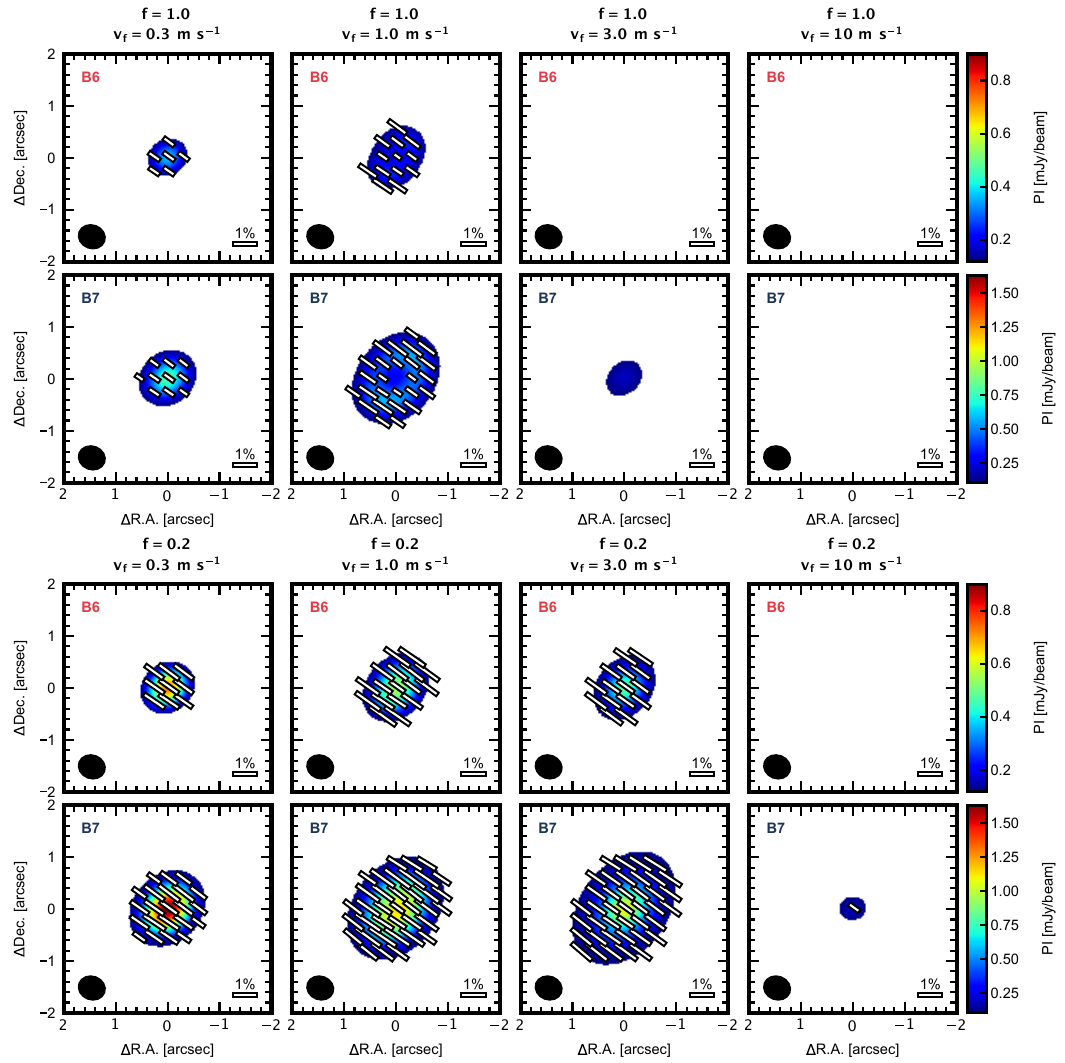}
\caption{
Full suite of polarization images obtained from our model.
The top two rows show the compact dust models, while the bottom two rows exhibit porous dust models.
The critical fragmentation velocity is 0.3, 1, 3 and 10 ${\rm m~s^{-1}}$ from left to right.
}
\label{fig:polarization_all}
\end{figure*}

\begin{figure*}[ht]
\centering
\includegraphics[width=1.0\textwidth]{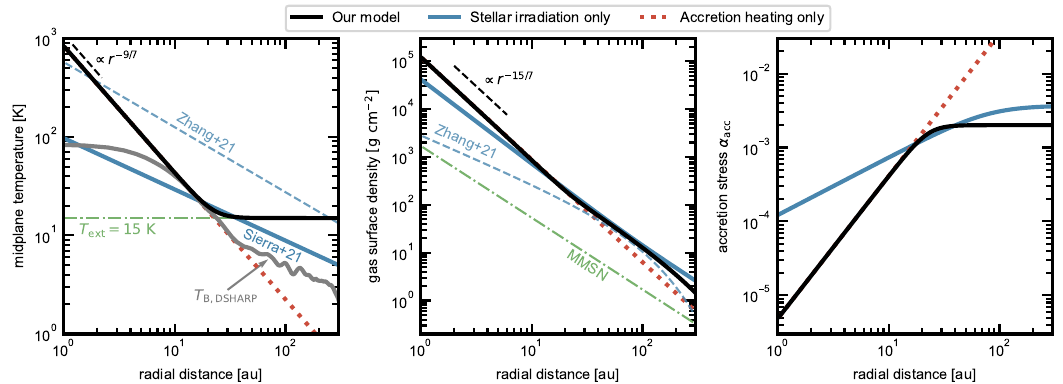}
\caption{
Summary of our model of the gas disk around IM Lup.
The black solid line denotes our model.
The blue solid line and red dotted line denotes the model with temperature solely determined by stellar irradiation ($T_{\rm MAPS}$; Eq. \ref{eq:Sierra}) and by accretion heating ($T_{\rm GSR,acc}$; Eq. \ref{eq:T_acc}), respectively.
The blue dashed lines show the conventional model \cite{Zhang+21} for which we assume $h_{\rm g}=c_{\rm s}/\Omega_{\rm K}$.
Left: midplane temperature profile. The gray solid line in the left panel denotes the observed brightness temperature at $\lambda=1.25~{\rm mm}$ taken by the ALMA DSHARP program \cite{Andrews+18}.
The green dash-dotted line denotes the temperature determined by external radiation (15 K).
Middle: gas surface density. The green dash-dotted line denotes the Minimum-Mass Solar Nebula model \cite{Hayashi81}.
Left: accretion parameter $\alpha_{\rm acc}$.
}
\label{fig:disk}
\end{figure*}

\begin{figure*}[ht]
\centering
\includegraphics[width=1.0\textwidth]{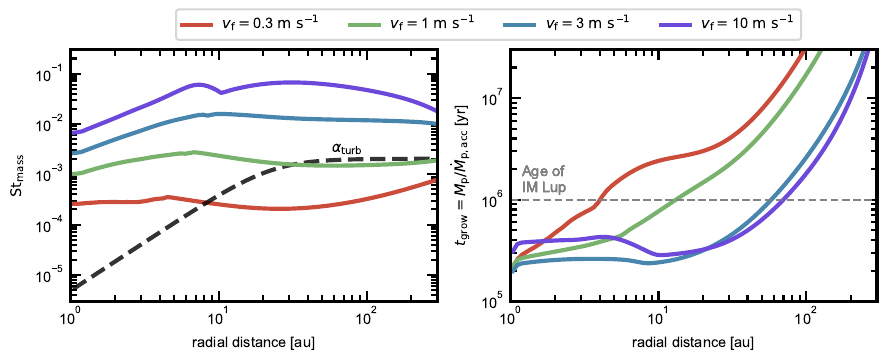}
\caption{
Representative dust size and core-growth timescale via pebble accretion estimated from full dust-size distiribution.
Left: mass-weighted average Stokes number. The black dashed line denotes the turbulence strength parameter $\alpha_{\rm turb}$ obtained from our model.
Right: core-growth timescale via pebble accretion. The horizontal dashed line denotes the typical estimate of the age of IM Lup (1 Myr \cite{Avenhaus+18}).
The core mass is assumed to be $3M_{\oplus}$.
}
\label{fig:pebble}
\end{figure*}

\clearpage
\subsection*{Parameter uncertainties}
Our model includes several parameters, for which we have fixed to the commonly adopted values for simplicity. Below, we briefly discuss the impact of their uncertainty on our conclusions.

\subsubsection*{Grain composition}
The dust composition is highly uncertain and affects the interpretation of the observations. Particularly, the inclusion of absorbing material (such as amorphous carbon \cite{Zubko+96}) increases the absorption opacity compared to the DSHARP model, thereby helping the model explain the observed bright millimeter emission. 
Detailed modeling of the NIR scattered light at outer region ($\gtrsim200$ au) of the IM Lup disk suggested that small grains responsible for the NIR scattered light are likely rich in amorphous carbon rather than organics \cite{Tazaki+23}. 
This is expected because amorphous carbon is produced by UV photodissociation of organics on the dust surface. 
However, it is highly uncertain whether large dust located deep in the disk midplane has the same composition as the small grains at the disk surface. 
From the point of view of millimeter polarization, the inclusion of amorphous carbon (i.e., absorbing material) makes scattering-induced polarization less efficient, which may contradict the observed high polarization degree. Based on these, we have chosen to use the commonly used DSHARP model for simplicity.

\subsubsection*{Ratio between $\alpha_{\rm acc}$, $\alpha_{\rm turb}$ and $\alpha_{\rm z}$}
We assume $\alpha_{\rm turb}=\alpha_{\rm acc}$ and $\alpha_{\rm z}=\alpha_{\rm acc}/3$.
If $\alpha_{\rm turb}$ is smaller than our assumed value, dust needs to be even more fragile to explain the observations. 
If $\alpha_{\rm turb}$ is larger, we expect $\alpha_{\rm acc}$ should be also higher because $\alpha_{\rm turb}$ represents the turbulence strength at the midplane where most of the disk mass resides. 
However, given the uncertainty of the measured mass accretion rate (factor of 2), $\alpha_{\rm acc}$ can be higher only by a factor of 2, which do not significantly affect our conclusion.
Constraining $\alpha_{\rm z}$ is even more difficult. 
We expect much smaller $\alpha_{\rm z}$ makes the model more deviates from the observations in terms of the following two aspects. 
First, more efficient dust settling makes accretion heating less efficient, which leads to less emission within 20 au. 
Secondly, more efficient dust settling makes the NIR scattering surface lower, which may contradict the NIR observation. 
These effects are sensitive to the distribution of small grains ($\lesssim10~{\rm \mu m}$) whose vertical distribution is affected only when ${\rm St}_{\rm 10{\mu m}}/\alpha_{\rm z}\gg1$. 
Our model shows ${\rm St}_{\rm 10{\mu m}}/\alpha_{\rm z}\sim1/7$ at 200 au,
and hence, the NIR surface would not be affected unless the actual $\alpha_{\rm z}$ is more than 7 times smaller than our model. 
Similar argument could be applied to the inner region; ${\rm St}_{\rm 10{\mu m}}/\alpha_{\rm z}\sim0.003$ at 10 au, suggesting that the midplane temperature at 10 au would not be affected even if $\alpha_{\rm z}$ is reduced by a factor of 100.

\subsubsection*{Initial dust-to-gas mass ratio}
The smaller dust-to-gas mass ratio makes the model disk fainter, while the higher dust mass helps the model explain the observed bright millimeter emission. 
If the dust-to-gas-mass ratio is 10 times higher than the fiducial value of 0.01, the $v_{\rm f}=10~{\rm m~s^{-1}}$ may produce the millimeter Stoke I emission similar to the observations. However, the $v_{\rm f}=10~{\rm m~s^{-1}}$ model already shows too large dust size to explain the millimeter polarization. 
Increasing the dust-to-gas mass ratio makes the situation even worse by enhancing the dust growth.

\subsubsection*{Mass accretion rate}
We assume the mass accretion rate of $10^{-7.9}M_{\odot}~{\rm yr^{-1}}$ and it has an uncertainty of factor of $\sim2$  \cite{Alcala+17}. 
This results in a factor of 2 uncertainty in $\alpha_{\rm acc}$ and hence $\alpha_{\rm turb}$. 
In the fragmentation-limited regime, the maximum dust size is proportional to $\alpha_{\rm turb}^{-1}v_{\rm f}^{2}$  \cite{Birnstiel+12}. 
Therefore, the factor of 2 uncertainty in the mass accretion rate propagates to the preferred $v_{\rm f}$ by only a factor of $\sim1.4$, which do not significantly affect our conclusion.

\subsubsection*{External heating}
Our assumed value (15 K) corresponds to the typical temperature of the Galactic molecular clouds \cite{Planck16}.
As both the surface density ($\Sigma_{\rm g} \propto T^{0.5}$) and black body radiation ($B_{\nu}(T) \propto T$ in the Rayleigh-Jeans limit) decreases with the temperature, lower $T_{\rm ext}$ makes it more difficult for the model to reproduce the observed bright thermal emission. 
Higher $T_{\rm ext}$ enhances the disk emission at outer region, which is helpful to explain the bright millimeter emission. 
However, the ALMA observations of the CO line emission suggests that it is likely that CO is (at least partially) freezed out at the disk midplane beyond 100 au, pointing to the temperature below the CO freeze-out temperature, $\sim20$ K \cite{Pinte+18,Zhang+23}. 
Based on these, we adopt a standard value of 15 K.

\renewcommand{\refname}{References in Supplementary information}



\begin{thebibliography}{1}
\bibitem{KL16} 
Kratter, K. \& Lodato, G., Gravitational Instabilities in Circumstellar Disks. ARAA 54, 271--311 (2016).

\bibitem{Perez+16} 
P{\'e}rez, L.~M. et al. Spiral density waves in a young protoplanetary disk. Science 353, 1519--1521 (2016).

\bibitem{Andrews+18} 
Andrews, S.~M. et~al. The Disk Substructures at High Angular Resolution Project (DSHARP). I. Motivation, Sample, Calibration, and Overview. ApJL 869, L41 (2018).

\bibitem{Tsukamoto+15} 
Tsukamoto, Y., Takahashi, S.~Z., Machida, M.~N. \& Inutsuka, S. Effects of radiative transfer on the structure of self-gravitating discs, their fragmentation and the evolution of the fragments. MNRAS 446, 1175-1190 (2015).

\bibitem{Helled+14}
Helled, R. et al. Giant Planet Formation, Evolution, and Internal Structure. Protostars and Planets VI, 643--665 (2014).

\bibitem{Alcala+17}
Alcal{\'a}, J.~M. et al. X-shooter spectroscopy of young stellar objects in Lupus. Accretion properties of class II and transitional objects. A\&A 600, A20 (2017).

\bibitem{Avenhaus+18}
Avenhaus, H. et al. Disks around T Tauri Stars with SPHERE (DARTTS-S). I. SPHERE/IRDIS Polarimetric Imaging of Eight Prominent T Tauri Disks. ApJ 863, 44 (2018).

\bibitem{Huang+18b}
Huang, J. et al. The Disk Substructures at High Angular Resolution Project (DSHARP). III. Spiral Structures in the Millimeter Continuum of the Elias 27, IM Lup, and WaOph 6 Disks. ApJL 869, L43 (2018).

\bibitem{Pinte+20}
Pinte , C. et al. Nine Localized Deviations from Keplerian Rotation in the DSHARP Circumstellar Disks: Kinematic Evidence for Protoplanets Carving the Gaps. ApJL 890, L9 (2020).

\bibitem{Tazaki+23}
Tazaki, R., Ginski, C. \& Dominik, C. Fractal Aggregates of Submicron-sized Grains in the Young Planet-forming Disk around IM Lup. ApJL 944, L43 (2023).

\bibitem{Ueda+19}
Ueda, T., Flock, M. \& Okuzumi, S. Dust Pileup at the Dead-zone Inner Edge and Implications for the Disk Shadow. ApJ 871, 10 (2019).

\bibitem{Sierra+21}
Sierra, A. et al. Molecules with ALMA at Planet-forming Scales (MAPS). XIV. Revealing Disk Substructures in Multiwavelength Continuum Emission. ApJS 257, 14 (2021).

\bibitem{Jiang+24}
Jiang, H., Mac{\'\i}as, E., Guerra-Alvarado, O.~M. \& Carrasco-Gonz{\'a}lez, C. Grain-size measurements in protoplanetary disks indicate fragile pebbles and low turbulence. A\&A 682, A32 (2024).

\bibitem{Birnstiel+12}
Birnstiel, T., Klahr, H. \& Ercolano, B. A simple model for the evolution of the dust population in protoplanetary disks. A\&A 539, A148 (2012).

\bibitem{Stephens+20}
Stephens, I. W. et al. Low-level Carbon Monoxide Line Polarization in Two Protoplanetary Disks: HD 142527 and IM Lup. ApJ 901, 71 (2020).

\bibitem{Hull+18}
Hull, C. L. H. et al. ALMA Observations of Polarization from Dust Scattering in the IM Lup Protoplanetary Disk. ApJ 860, 82 (2018).

\bibitem{Tazaki+19}
Tazaki, R. et al. Unveiling Dust Aggregate Structure in Protoplanetary Disks by Millimeter-wave Scattering Polarization. ApJ 885, 52 (2019).

\bibitem{Zhang+23}
Zhang, S. et al. Porous Dust Particles in Protoplanetary Disks: Application to the HL Tau Disk. ApJ 953, 96 (2023).

\bibitem{OT19}
Okuzumi, S. \& Tazaki, R. Nonsticky Ice at the Origin of the Uniformly Polarized Submillimeter Emission from the HL Tau Disk. ApJ 878, 132 (2019).

\bibitem{Zhang+21}
Zhang, K. et al. Molecules with ALMA at Planet-forming Scales (MAPS). V. CO Gas Distributions. ApJS 257, 5 (2021).

\bibitem{Law+21}
Law, C. J. et al. Molecules with ALMA at Planet-forming Scales (MAPS). IV. Emission Surfaces and Vertical Distribution of Molecules. ApJS 257, 4 (2021).

\bibitem{Dominik+16}
Dominik, C., Paszun, D. \& Borel, H.
The structure of dust aggregates in hierarchical coagulation. arXiv e-prints arXiv:1611.00167 (2016).

\bibitem{Tanaka+23}
Tanaka, H., Anayama, R. \& Tazaki, R. Compression of Dust Aggregates via Sequential Collisions with High Mass Ratios. ApJ 945, 68 (2023).

\bibitem{Huang+18}
Huang, J. et al. The Disk Substructures at High Angular Resolution Project (DSHARP). II. Characteristics of Annular Substructures. ApJL 869, L42 (2018).

\bibitem{Verrios+22}
Verrios, H. J. et al. Kinematic Evidence for an Embedded Planet in the IM Lupi Disk. ApJL 934, L11 (2022).

\bibitem{Dullemond+18}
Dullemond, C.~P. et al. The Disk Substructures at High Angular Resolution Project (DSHARP). VI. Dust Trapping in Thin-ringed Protoplanetary Disks. ApJL 869, L46 (2018).

\bibitem{LC18}
Liu, B. \& Ormel, C. W. Catching drifting pebbles. I. Enhanced pebble accretion efficiencies for eccentric planets. A\&A 615, A138 (2018).

\bibitem{Lau+22}
Lau, T. C. H. et al. Rapid formation of massive planetary cores in a pressure bump. A\&A 668, A170 (2022).

\bibitem{JO+23}
Jiang, H. \& Ormel, C. W. Efficient planet formation by pebble accretion in ALMA rings. MNRAS 518, 3877-3900 (2023).

\bibitem{Crida+09}
Crida, A., Masset, F. \& Morbidelli, A.
Long Range Outward Migration of Giant Planets, with Application to Fomalhaut b. ApJL 705, L148--L152 (2009).

\bibitem{Boss97}
Boss, A. P. Giant planet formation by gravitational instability. Science 276, 1836--1839 (1997).

\bibitem{Ormel17}
Ormel, C. W. The Emerging Paradigm of Pebble Accretion. Formation, Evolution, and Dynamics of Young Solar Systems, 445, 197 (2017).

\bibitem{BH91}
Balbus, S. A. \& Hawley, J. F. A Powerful Local Shear Instability in Weakly Magnetized Disks. I. Linear Analysis. ApJ 376, 214 (1991).

\bibitem{Bai17}
Bai, X.-N. Global Simulations of the Inner Regions of Protoplanetary Disks with Comprehensive Disk Microphysics. ApJ 845, 75 (2017).

\bibitem{Mori+19}
Mori, S., Bai, X.-N. \& Okuzumi, S. Temperature Structure in the Inner Regions of Protoplanetary Disks: Inefficient Accretion Heating Controlled by Nonideal Magnetohydrodynamics. ApJ 872, 98 (2019).

\bibitem{BL20}
B{\'e}thune, W. \& Latter, H. Electric heating and angular momentum transport in laminar models of protoplanetary discs. MNRAS 494, 6103--6119 (2020).

\bibitem{Kondo+23}
Kondo, K., Okuzumi, S. \& Mori, S. The Roles of Dust Growth in the Temperature Evolution and Snow Line Migration in Magnetically Accreting Protoplanetary Disks. ApJ 949, 119 (2023).

\bibitem{OU21}
Ohno, K. \& Ueda, T.
Jupiter's ``cold'' formation in the protosolar disk shadow. An explanation for the planet's uniformly enriched atmosphere. A\&A 651, L2 (2021).

\bibitem{Notsu+22}
Notsu, S. et al. The Molecular Composition of Shadowed Proto-solar Disk Midplanes Beyond the Water Snowline. ApJ 936, 188 (2022).

\bibitem{Seifert+21}
Seifert, R. A., et al. Evidence for a Cosmic-Ray Gradient in the IM Lup Protoplanetary Disk. ApJ 912, 136 (2021).

\bibitem{Mannella+96}
Mennella, V. et al. Activation of an Ultraviolet Resonance in Hydrogenated Amorphous Carbon Grains by Exposure to Ultraviolet Radiation. ApJL 464, L191 (1996).

\bibitem{Bergin+07}
Bergin, E. A. et al. The Chemical Evolution of Protoplanetary Disks. Protostars and Planets V, 751--766 (2007).

\bibitem{Bosman+21}
Bosman, A. D. et al. Molecules with ALMA at Planet-forming Scales (MAPS). XV. Tracing Protoplanetary Disk Structure within 20 au. ApJS 257, 15 (2021).

\bibitem{Bosman+23}
Bosman, A. D. et al. A Potential Site for Wide-orbit Giant Planet Formation in the IM Lup Disk. ApJL 944, L53 (2023).

\end{thebibliography}

\begin{thebibliography}{1}
\setcounter{enumiv}{44}

\bibitem{Paczynski78}
Paczynski, B. A model of selfgravitating accretion disk.
Acta Astron. 28, 91--109 (1978).

\bibitem{Gammie01}
Gammie, C. F. Nonlinear Outcome of Gravitational Instability in Cooling, Gaseous Disks. ApJ 553, 174--183 (2001).

\bibitem{Tsukamoto+17}
Tsukamoto, Y., Okuzumi, S. \& Kataoka, A. Apparent Disk-mass Reduction and Planetisimal Formation in Gravitationally Unstable Disks in Class 0/I Young Stellar Objects. ApJ 838, 151 (2017).

\bibitem{Yamamuro+23}
Yamamuro, R., Tanaka, K. E. I. \& Okuzumi, S. Massive Protostellar Disks as a Hot Laboratory of Silicate Grain Evolution. ApJ 949, 29 (2023).

\bibitem{XA23}
Xu, W. \& Armitage, P. J. Revisiting Collisional Dust Growth in Class 0/I Protostellar Disks: Sweep-up Can Convert a Few 10 M $_{{\ensuremath{\oplus}}}$ of Dust into Kilogram Pebbles in 0.1 Myr. ApJ 946, 94 (2023).

\bibitem{Toomre1964}
Toomre, A. On the gravitational stability of a disk of stars. ApJ 139, 1217--1238 (1964).

\bibitem{LD94}
Laughlin, G. \& Bodenheimer, P. Nonaxisymmetric Evolution in Protostellar Disks. ApJ 436, 335 (1994).

\bibitem{SS73}
Shakura, N. I. \& Sunyaev, R. A. Black holes in binary systems. Observational appearance. A\&A 24, 337-355 (1973).

\bibitem{Oberg+21}
\"{O}berg, K.~I. et al. Molecules with ALMA at Planet-forming Scales (MAPS). I. Program Overview and Highlights. ApJS 257, 1 (2021).

\bibitem{XK21b}
Xu, W. \& Kunz, M. W. Formation and evolution of protostellar accretion discs - II. From 3D simulation to a simple semi-analytic model of Class 0/I discs. MNRAS 508, 2142--2168 (2021).

\bibitem{Cleeves+16}
Cleeves, L. I. et al. The Coupled Physical Structure of Gas and Dust in the IM Lup Protoplanetary Disk. ApJ 832, 110 (2016).

\bibitem{Powell+22}
Powell, D. et al. Depletion of gaseous CO in protoplanetary disks by surface-energy-regulated ice formation. Nature Astronomy 6, 1147--1155 (2022).

\bibitem{Lodato+23}
Lodato, G. et al. Dynamical mass measurements of two protoplanetary discs. MNRAS 518, 4481--4493 (2023).

\bibitem{Martire+24}
Martire, P. et al. Rotation curves in protoplanetary disks with thermal stratification. arXiv e-prints arXiv:2402.12236 (2024).

\bibitem{Franceschi+23}
Franceschi, R., et al. Constraining the turbulence and the dust disk in IM Lup: Onset of planetesimal formation. A\&A 671, A125 (2023).

\bibitem{Rosotti23}
Rosotti, G. P. Empirical constraints on turbulence in proto-planetary discs. New Astronomy Reviews 96, 101674 (2023).

\bibitem{Paneque+24}
Paneque-Carre{\~n}o, T. et al. High turbulence in the IM Lup protoplanetary disk: Direct observational constraints from CN and C$_2$H emission. A\&A 684, A174 (2024).

\bibitem{Rice+11}
Rice, W. K. M. et al. Stability of self-gravitating discs under irradiation. MNRAS 418, 1356--1362 (2011).

\bibitem{Paardekooper12}
Paardekooper, S.-J. Numerical convergence in self-gravitating shearing sheet simulations and the stochastic nature of disc fragmentation. MNRAS 421, 3286-3299 (2012).

\bibitem{Bethune+21}
B{\'e}thune, W., Latter, H. \& Kley, W. Spiral structures in gravito-turbulent gaseous disks. A\&A 650, A49 (2021).

\bibitem{Stammler+22}
Stammler, S. M. \& Birnstiel, T. DustPy: A Python Package for Dust Evolution in Protoplanetary Disks. ApJ 935, 35 (2022).

\bibitem{Birnstiel+11}
Birnstiel, T., Ormel, C. W. \& Dullemond, C. P. Dust size distributions in coagulation/fragmentation equilibrium: numerical solutions and analytical fits. A\&A 525, A11 (2011).

\bibitem{OC07}
Ormel, C. W. \& Cuzzi, J. N. Closed-form expressions for particle relative velocities induced by turbulence. A\&A 466, 413--420 (2007).

\bibitem{Beitz+11}
Beitz, E. et al. Low-velocity Collisions of Centimeter-sized Dust Aggregates. ApJ 736, 34 (2011).

\bibitem{Wada+13}
Wada, K. et al. Growth efficiency of dust aggregates through collisions with high mass ratios. A\&A 559, A62 (2013).

\bibitem{Arakawa+16}
Arakawa, S. \& Nakamoto, T. Rocky Planetesimal Formation via Fluffy Aggregates of Nanograins. ApJL 832, L19 (2016).

\bibitem{Okuzumi+16}
Okuzumi, S. et al. Sintering-induced Dust Ring Formation in Protoplanetary Disks: Application to the HL Tau Disk. ApJ 821, 82 (2016).

\bibitem{MW19}
Musiolik, G. \& Wurm, G. Contacts of Water Ice in Protoplanetary Disks{\textemdash}Laboratory Experiments. ApJ 873, 58 (2019).

\bibitem{Musiolik+16a}
Musiolik, G. et al. Collisions of CO$_{2}$ Ice Grains in Planet Formation. ApJ 818, 16 (2016).

\bibitem{Musiolik+16b}
Musiolik, G. et al. Ice Grain Collisions in Comparison: CO2, H2O, and Their Mixtures. ApJ 827, 63 (2016).

\bibitem{RADMC}
Dullemond, C. P. et al. RADMC-3D: A multi-purpose radiative transfer tool. Astrophysics Source Code Library, record ascl:1202.015 (2012).

\bibitem{Birnstiel+18}
Birnstiel, T. et al. The Disk Substructures at High Angular Resolution Project (DSHARP). V. Interpreting ALMA Maps of Protoplanetary Disks in Terms of a Dust Model. ApJ 869, L45 (2018).

\bibitem{WB08}
Warren, S. G. \& Brandt, R. E. Optical constants of ice from the ultraviolet to the microwave: A revised compilation. Journal of Geophysical Research (Atmospheres) 113, D14220 (2008).

\bibitem{Draine03}
Draine, B. T. Interstellar Dust Grains. ARAA 41, 241--289 (2003).

\bibitem{HS96}
Henning, T. \& Stognienko, R.
Dust opacities for protoplanetary accretion disks: influence of dust aggregates. A\&A 311, 291-303 (1996).

\bibitem{Optool}
Dominik, C., Min, M. \& Tazaki, R. OpTool: Command-line driven tool for creating complex dust opacities. Astrophysics Source Code Library, record ascl:2104.010 (2021).

\bibitem{Zubko+96}
Zubko, V. G. et al. Optical constants of cosmic carbon analogue grains - I. Simulation of clustering by a modified continuous distribution of ellipsoids. MNRAS 282, 1321--1329 (1996).

\bibitem{Min+05}
Min, M., Hovenier, J. W. \& de Koter, A. Modeling optical properties of cosmic dust grains using a distribution of hollow spheres. A\&A 432, 909--920 (2005).

\bibitem{Kataoka+14}
Kataoka, A. et al. Opacity of fluffy dust aggregates. A\&A 568, A42 (2014).

\bibitem{Kataoka+15}
Kataoka, A. et al. Millimeter-wave Polarization of Protoplanetary Disks due to Dust Scattering. ApJ 809, 78 (2015).

\bibitem{Woitke+23}
Woitke, P. et al. 2D disc modelling of the JWST line spectrum of EX Lupi. A\&A 683, A219 (2024).

\bibitem{FP06}
Fromang, S. \& Papaloizou, J. Dust settling in local simulations of turbulent protoplanetary disks. A\&A 452, 751--762 (2006).

\bibitem{YL07}
Youdin, A. N. \& Lithwick, Y. Particle stirring in turbulent gas disks: Including orbital oscillations. Icarus 192, 588--604 (2007).

\bibitem{HG41}
Henyey, L. G. \& Greenstein, J. L. Diffuse radiation in the Galaxy. ApJ 93, 70--83 (1941).

\bibitem{Pinte+18}
Pinte, C. et al. Direct mapping of the temperature and velocity gradients in discs. Imaging the vertical CO snow line around IM Lupi. A\&A 609, A47 (2018).

\end{thebibliography}

\begin{thebibliography}{1}

\bibitem{Hayashi81}
Hayashi, C. Structure of the Solar Nebula, Growth and Decay of Magnetic Fields and Effects of Magnetic and Turbulent Viscosities on the Nebula. Progress of Theoretical Physics Supplement, 70, 35--53 (1981).

\bibitem{Planck16}
Planck Collaboration et al. Planck 2015 results. XXVIII. The Planck Catalogue of Galactic cold clumps. A\&A 594, A28 (2016).

\end{thebibliography}
\end{document}